\documentstyle[plenum,epsfig,12pt]{book}
\begin{document}
\newcommand{\be}{\begin{eqnarray}}
\newcommand{\ee}{\end{eqnarray}}

\chapter{UNIVERSAL BEHAVIOR IN DIRAC SPECTRA}

\author{Jacobus Verbaarschot}

\affiliation{Department of Physics and Astronomy\\
University at Stony Brook, SUNY\\
Stony Brook, NY 11794, USA}
 
\section{INTRODUCTION}

In the chiral limit the QCD Lagrangian has an important global chiral 
symmetry: Specifically, for QCD with three colors and $N_f$ massless flavors, 
the invariance group acting on the quark fields 
is $SU(N_f) \times SU(N_f) \times U_A(1)$. In spite of the lightness of the
up and down quark, no signatures of this symmetry have been observed in nature.
On the contrary, the fact that the pion mass is much lighter than the other
hadron masses indicates that $SU(N_f) \times SU(N_f)$
is spontaneous broken with the pions
as Goldstone bosons. The breaking of $U_A(1)$ follows from the
absence of parity doublets. For example, the mass of the $\delta$ is very 
different from the mass of the pion. The description of 
the broken phase requires the introduction of an order parameter 
which is zero in the restored phase.
For the chiral phase transition the order parameter is the 
chiral condensate $\langle \bar \psi \psi \rangle$. 
Large scale lattice QCD simulations performed during the past decade show that
for two light flavors the value of the 
chiral condensate is approximately 
$-(220\, MeV)^3$ at zero temperature and vanishes above
a critical temperature of about 140 $MeV$
(see reviews by
DeTar\refnote{\cite{DeTar}}, Ukawa\refnote{\cite{Ukawa}} and 
Smilga\refnote{\cite{Smilref}} for recent results on this topic).
As was
in particular pointed out by Shuryak\refnote{\cite{Edwardaxial}}, not
necessarily both symmetries are restored at the same 
temperature\refnote{\cite{DeTarU1}}. This may  lead
to interesting observable consequences.  

According the Banks-Casher formula\refnote{\cite{Banks-Casher}}, 
the chiral condensate is directly 
related to the spectrum of the Euclidean Dirac operator near zero virtuality
(here and below, we use virtuality for the value 
of the Euclidean Dirac eigenvalues).
However, the eigenvalues $fluctuate$ about their average position as the
gauge fields vary over the ensemble of gauge field configurations.
The main question that will be addressed in these two lectures 
is to what extent such fluctuations
are universal. Inspired by the study of 
spectra of complex systems\refnote{\cite{bohigas}},
we will conjecture that spectral fluctuations on the scale
of a typical eigenvalue spacing are universal\refnote{\cite{SVR}}. 
If this  conjecture is true, such spectral fluctuations can be obtained 
from a broad class of theories with global symmetries of QCD as common
ingredient. We will derive them from the simplest models in this class:
chiral Random Matrix Theory (chRMT).

A first argument in favor of 
universality in Dirac spectra
came from the analysis of the finite volume QCD partition 
function\refnote{\cite{LS}}.
As has been shown by Gasser and Leutwyler\refnote{\cite{GL}},
for box size $L$ in the range 
\be
1/\Lambda \ll L \ll 1/m_\pi,
\label{range}
\ee
($\Lambda$ is a typical hadronic scale and $m_\pi$ is the pion mass)
the mass dependence of the QCD partition function is completely
determined by its global symmetries. As a consequence, fluctuations of 
Dirac eigenvalues near zero virtuality are constrained by,
but not determined by, an infinite family of sum 
rules\refnote{\cite{LS}} (also called Leutwyler-Smilga sum rules). 
For example, the simplest Leutwyler-Smilga sum rule can be obtained from the
microscopic spectral density\refnote{\cite{SVR}} 
(the spectral density near zero virtuality on 
a scale of a typical eigenvalue
spacing). On the other hand, the infinite family of Leutwyler-Smilga
sum rules is not sufficient to determine the microscopic spectral density.
The additional ingredient is
universality. A priori there is no reason that fluctuations of 
Dirac eigenvalues are in the same universality class as chRMT. Whether or not
QCD is inside this class is a dynamical question that can only be answered
by full scale lattice QCD simulations. However, the confidence in an 
affirmative answer to this question has been greatly enhanced by 
universality studies within chiral Random Matrix Theory. 
The aim of such studies is to show that
spectral fluctuations do not depend on the details of the probability 
distribution. Recently, it has been shown that the microscopic spectral
density is universal for a wide class of probability 
distributions\refnote{\cite{Nishigaki-uni,
Damgaard,brezin-hikami-zee,GWu,Sener1,Seneru}}. 
We will give an extensive review of these important new results.

Because of the $U_A(1)$ symmetry of the Dirac operator two types
of spectral fluctuations can be distinguished. Spectral fluctuations 
near zero virtuality and spectral fluctuations in the bulk of the spectrum.
The fluctuations of Dirac eigenvalues near zero virtuality
are directly related to the approach to the thermodynamic limit of
the chiral condensate. In particular, knowledge of the microscopic 
spectral density provides us with a quantitative 
explanation\refnote{\cite{VPLB}} of
finite size corrections to the valence quark mass of 
dependence of the chiral condensate\refnote{\cite{Christ}}. 

Recently, it has become possible to obtain $all$ eigenvalues of the lattice
QCD Dirac operator on reasonably large 
lattices\refnote{\cite{Kalkreuter,berbenni}}, making 
a direct verification of the above conjecture possible. 
This is one of the main objectives of these lectures. This is easiest
for correlations in the bulk of the spectrum. Under the assumption
of spectral ergodicity\refnote{\cite{pandey}}, 
eigenvalue correlations can be studied by spectral 
averaging instead of ensemble averaging\refnote{\cite{HV,HKV}}. 
On the other hand, in order to 
study the microscopic spectral density, a very large number of independent
gauge field configurations is required. First lattice results confirming
the universality of the microscopic spectral density have been obtained
recently\refnote{\cite{berbenni}}.

At this point I wish to stress that there are two different types of 
applications
of Random Matrix Theory. In the first type, fluctuations
of an observable are related to its average. Because of universality
it is possible to obtain exact results. In general, the average of 
an observable is not given by Random Matrix Theory. There are many examples
of this type of universal fluctuations ranging
from atomic physics to quantum field theory (a recent comprehensive review
was written by Guhr, M\"uller-Groeling and
Weidenm\"uller\refnote{\cite{hdgang}}). Most examples are related to
fluctuations of eigenvalues. Typical examples are
nuclear spectra\refnote{\cite{Haq}},
acoustic spectra\refnote{\cite{Guhr}}, resonances in resonance 
cavities\refnote{\cite{Koch}},
$S$-matrix fluctuations\refnote{\cite{ERICSON,WEIDI}} and
universal conductance fluctuations\refnote{\cite{meso}}.
In these lectures we will discuss correlations in the bulk of Dirac spectra and
the microscopic spectral density.
The second type of application of Random Matrix Theory is as a schematic
model of disorder.  In this way one obtains $qualitative$ results which
may be helpful in understanding some physical phenomena. There are numerous
examples in this category. We only mention the Anderson 
model of localization\refnote{\cite{Anderson}}, neural 
networks\refnote{\cite{Sommers}},
the Gross-Witten model of QCD\refnote{\cite{GW}} 
and quantum gravity\refnote{\cite{Ginsparg}} . In these 
lectures we will discuss chiral random matrix models at nonzero 
temperature and chemical
potential. In particular, we will review recent work
by Stephanov\refnote{\cite{Stephanov}} on the quenched approximation at
nonzero chemical potential.

In the first lecture we will review some general properties of Dirac spectra
including the Banks-Casher formula. From the zeros of the partition function
we will show that there is an intimate relation between chiral symmetry
breaking and correlations of Dirac eigenvalues. Starting from Leutwyler-Smilga
sum-rules the microscopic spectral density will be introduced. We will discuss
the statistical analysis of quantum spectra. It will be argued
that  spectral correlations of 'complex' systems 
are given by Random Matrix Theory.
We will end the first lecture with the introduction of 
chiral Random Matrix Theory.

In the second lecture we will compare the chiral random matrix model with 
QCD and discuss some of its properties. We will review recent
results showing that the microscopic
spectral density and eigenvalue correlations near zero virtuality are 
strongly universal. 
Lattice QCD results for the microscopic spectral
density and correlations in the bulk of the spectrum will be discussed in
detail. We will end the second lecture with a review  of chiral Random
Matrix Theory at nonzero chemical potential. Novel features of spectral
universality in nonhermitean matrices will be discussed.
 
\section{THE DIRAC SPECTRUM}
\subsection{Introduction}
The Euclidean QCD partition function is given by
\be
Z_{\rm QCD }(m,\theta) = 
\int dA \det{(\gamma D +m)} e^{-S_{\rm YM}/\hbar+i\theta\nu},
\label{QCDpart}
\ee
where $\gamma D=\gamma_\mu (\partial_\mu + i A_\mu)$ 
is the anti-Hermitean Dirac operator and $S_{\rm YM}$ is
the Yang-Mills action. The integral over field configurations includes a
sum over all topological sectors with topological charge $\nu$. Each sector
is weighted by $\exp(i\theta \nu)$. 
Phenomenologically the value of the vacuum
$\theta$-angle is consistent with zero.
We use the convention that the Euclidean gamma matrices are
Hermitean with $\{\gamma_\mu, \gamma_\nu\} = 2 \delta_{\mu\nu}$.
The integral is over all gauge field configurations,
and for definiteness, we assume a lattice regularization of the 
partition function. 

Our main object of interest is the spectrum of the Dirac operator. The
eigenvalues $\lambda_k$ are defined by
\be
\gamma D\phi_k = i\lambda_k \phi_k.
\ee
The spectral density is given by
\be
\rho(\lambda) = \sum_k \delta (\lambda-\lambda_k).
\ee
Correlations of the eigenvalues can be expressed in terms of 
the two-point correlation
function
\be
\rho_2(\lambda,\lambda') = \langle \rho(\lambda) \rho(\lambda') \rangle,
\ee
where $\langle \cdots \rangle$ denotes averaging with respect to the
QCD partition function (\ref{QCDpart}). The connected two-point
correlation function is obtained by subtraction of the product of the average
spectral densities
\be
\rho_c(\lambda,\lambda') = \rho_2(\lambda,\lambda')-
\langle \rho(\lambda)\rangle \langle \rho(\lambda') \rangle.
\ee

Because of the $U_A(1)$  symmetry
\be
\{\gamma_5, \gamma D\} = 0,
\ee
the eigenvalues occur in pairs $\pm \lambda$ or are zero. The eigenfunctions
are given by $\phi_k$ and $\gamma_5 \phi_k$, respectively. If $\gamma_5 \phi_k
= \pm \phi_k$, then necessarily $\lambda_k = 0$. This happens for a
solution of the Dirac operator in the field of an instanton. In
a sector with topological charge $\nu$ the Dirac operator has 
$\nu$ exact zero modes with
positive chirality. In order to represent the low energy sector of the 
Dirac operator for field configurations with topological charge $\nu$, 
it is natural to choose a chiral basis with 
$n$ right-handed states and $m\equiv n+\nu$ left-handed states. Then
the Dirac matrix has the block structure
\be
\left ( \begin{array}{cc} 0 & W \\ W^\dagger & 0 \end{array}\right ),
\label{block}
\ee
where $W$ is an $n\times m$ matrix. For $m=2$ and $n=1$, one can easily 
convince oneself that the Dirac matrix has exactly one zero eigenvalue. We 
leave it as an exercise to the reader to show that in general the Dirac matrix
has $|m-n|$ zero eigenvalues.

In terms of the eigenvalues of the Dirac operator, the QCD partition
function can be rewritten as
\be
Z_{\rm QCD }(m,\theta) =\sum_\nu e^{i\nu \theta}\prod_f m_f^{|\nu|} 
\int_{\nu} dA \prod_f\prod_{k}(\lambda_k^2 +m^2_f) e^{-S_{YM}/\hbar}
\label{ZQCDdet}
\ee
where $\int_\nu dA$ denotes the integral over field configurations with
topological charge $\nu$, and $\prod_f$ is the product
over $N_f$ flavors with mass $m_f$.
The partition function in the sector of topological charge $\nu$ is obtained
by Fourier inversion
\be
Z_\nu(m) = \frac 1{2\pi}\int_0^{2\pi} d\theta e^{-i\nu\theta} 
Z_{\rm QCD}(m,\theta).
\label{znum}
\ee 
The fluctuations of the eigenvalues of the QCD Dirac operator 
are induced by the fluctuations of the gauge fields. 
Formally, one can think of integrating out all gauge field configurations
for fixed values of the Dirac eigenvalues. 
The transformation of integration variables from the fields, $A$, to the
eigenvalues, $\lambda_k$,  leads to a nontrivial "Jacobian".
Universality in Dirac spectra has its origin in this "Jacobian".

The free Dirac spectrum can be obtained immediately from the square of the
Dirac operator. For a box of volume $L_1\times 
L_2\times L_3\times L_4$ one finds
\be
\lambda_{\vec n} = 2\pi\left (
(\frac {n_1}{L_1})^2+(\frac {n_2}{L_2})^2+
(\frac {n_3}{L_3})^2+(\frac {n_4+\frac 12}{L_4})^2\right)^{1/2},
\ee
where we have used periodic boundary conditions in the spatial directions
and anti-periodic boundary conditions in the time direction.
The spectral density is obtained by counting the total number of eigenvalues
in a shell of radius $\lambda L/2\pi$. The result is
\be
\rho^{\rm free}(\lambda) \sim V \lambda^3.
\label{rhofree}
\ee
For future reference, we note that in the generic case, when the sides of
the hypercube are related by an irrational number, asymptotically, 
the eigenvalues are  uncorrelated, i.e.
\be
\rho_2(\lambda, \lambda') = \langle \rho(\lambda)\rangle 
\langle \rho(\lambda')\rangle.
\ee

\begin{center}
\begin{figure}[!ht]
\vspace{0.5cm}
\centering\includegraphics[width=100mm,angle=0]{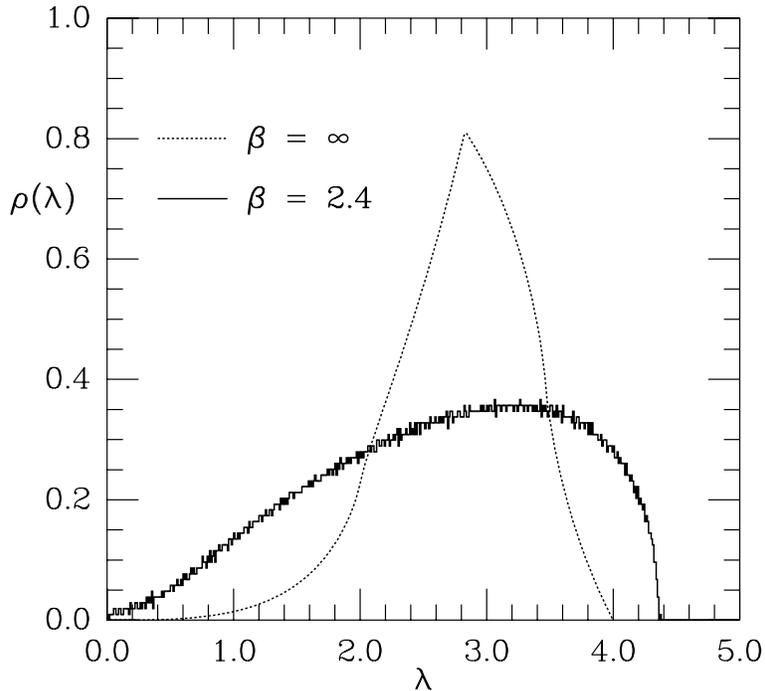}
\caption{The free Dirac spectral density (dotted curve) and the spectral
density of the Kogut-Susskind Dirac operator
for a gauge field configuration generated with $\beta = 2.4$
(histogram). Both spectral densities are on a $12^4$ lattice
and are normalized to unit area.} 
\label{fig1}
\vspace{-0.5cm}
\end{figure}
\end{center}

Two examples of Dirac spectra are shown in Fig. 1. The dotted curve 
represents the free Kogut-Susskind Dirac spectrum  on a $12^4$ lattice
with periodic boundary conditions in the spatial directions
anti-periodic boundary conditions in the time direction. For an
$N_1\times N_2\times N_3 \times N_4$ lattice this spectrum is given by
\be
\lambda_{\vec n} = 2\left ( 
\sin^2 \left ( \frac {\pi n_1}{N_1} \right ) +
\sin^2 \left ( \frac {\pi n_2}{N_2} \right ) +
\sin^2 \left ( \frac {\pi n_3}{N_3} \right ) +
\sin^2 \left ( \frac {\pi (n_4+0.5)}{N_4} \right ) \right)^{1/2}.
\ee
Here, $n_i = 0, 1, \cdots, [N_i/2]$ ($i = 1,\, 2, \,3$) and $n_4 = 0, 1, \cdots,
[(N_4+1)/2]$. The  Kogut-Susskind Dirac
spectrum for an $SU(2)$ gauge field configuration with $\beta =2.4$
on the same size lattice is shown by the histogram in the same figure (full 
curve). We clearly observe
an accumulation of small eigenvalues.

\subsection{The Banks-Casher Relation}

The order parameter of the chiral phase transition,
$\langle \bar \psi \psi \rangle$,
is nonzero only below the critical temperature. 
As was shown by Banks and Casher\refnote{\cite{Banks-Casher}}, 
$\langle \bar \psi \psi \rangle$ is directly related to the eigenvalue density
of the QCD Dirac operator per unit four-volume 
\be
\Sigma \equiv 
|\langle \bar \psi \psi \rangle| =\lim\frac {\pi \langle {\rho(0)}\rangle}V.
\label{bankscasher}
\ee
It is elementary to derive this relation.
The chiral condensate follows from
the partition function (\ref{ZQCDdet}) (all quark masses are chosen equal),
\be
\langle \bar \psi \psi \rangle  = -\lim\frac 1{VN_f}\partial_m \log Z(m) 
=-\lim\frac 1V \langle \sum_k \frac {2m}{\lambda_k^2 + m^2}\rangle.
\label{preBC}
\ee
If we express the sum as an integral over the average spectral density,
and take the thermodynamic limit before the chiral limit, so that many
eigenvalues are less than $m$, we recover (\ref{bankscasher}). The order of the
limits in (\ref{bankscasher}) is important. First we take the 
thermodynamic limit, next the chiral limit and, finally, the field theory
limit. As can be observed from (\ref{preBC}) the
sign of $\langle \bar \psi \psi \rangle $ changes if $m$ crosses the 
imaginary axis. 

An important consequence of the Bank-Casher formula (\ref{bankscasher})
is that the eigenvalues near zero virtuality are spaced as  
\be
\Delta \lambda = 1/{\rho(0)} = {\pi}/{\Sigma V}.
\ee
This should be contrasted with the eigenvalue spectrum 
of the non-interacting 
Dirac operator. Then eq. (\ref{rhofree}) results in
an eigenvalue spacing equal to  $\Delta \lambda \sim 1/V^{1/4}$.
Clearly, the presence of gauge fields leads to a strong modification of
the spectrum near zero virtuality. Strong interactions result in the 
coupling of many degrees of freedom leading to extended states and correlated
eigenvalues.
On the other hand, for uncorrelated eigenvalues, the eigenvalue distribution
factorizes, and for $\lambda \ne 0$,
we have $\rho(\lambda) \sim \lambda^{2N_f+1}$ in the chiral limit, 
i.e. no breaking of chiral symmetry. 
One consequence of the interactions is level repulsion
of neighboring eigenvalues. Therefore, the two smallest eigenvalues of
the Dirac operator, $\pm \lambda_{\rm min}$ repel each other, and the Dirac
spectrum will have a gap at $\lambda = 0$ with a width of the order of 
$1/\Sigma V$.  

\section{Spectral Correlations and Zeros of the Partition Function}

The study of zeros of the partition function has been a fruitful
tool in statistical mechanics\refnote{\cite{Frank,Shrock}}. In QCD,
both zeros in the complex fugacity plane and the complex mass
plane have been studied\refnote{\cite{vink,barbourqed}}.
Since the QCD partition function is a polynomial in $m$ it can be factorized
as (all quark masses are taken to be equal to $m$)
\be
Z_{\rm QCD}(m,\theta) = \prod_k (m-m_k).
\ee
Because configurations of opposite topological charge occur with the same 
probability,
the coefficients of this polynomial are real, and 
the zeros occur in complex conjugate pairs.  For an even number of flavors
the zeros occur in pairs $\pm m_k$. In a sector with topological charge
$\nu$, this is also the case for even
$N_f \times \nu$. The chiral condensate is given by
\be
\Sigma(m) = -\lim\frac 1{VN_f} \partial_m \log Z_{\rm QCD}(m,\theta)=
-\lim\frac 1{VN_f}\sum_k \frac 1{m-m_k}.
\ee
For an even number of flavors, $\Sigma(m)$ is an odd function of $m$.
In order to have a discontinuity at $m=0$, the zeros in this region
have to coalesce into a 
cut along the imaginary axis in the thermodynamic limit.

In the hypothetical case that the eigenvalues of the Dirac operator do not
fluctuate the zeros are located at $m_k = \pm i\lambda_k$. In the opposite
case, of uncorrelated eigenvalues, the eigenvalue distribution factorizes
and all zeros are located at $\pm i \sigma$, where $\sigma^2 = \langle
\lambda^2_k\rangle$.  
As a result, the chiral condensate does not 
show a discontinuity across the imaginary axis and is equal to zero.

We hope to convince the reader that the presence of a  
discontinuity is intimately related to  correlations of 
eigenvalues\refnote{\cite{oscor}}. 
Let us study the effect of pair correlations for one flavor in 
the sector of zero topological charge. 
The fermion determinant can be written as
\be
\langle \prod_k (m^2+\lambda_k^2) \rangle = \sum_k 
\left (\begin{array}{c}  N \\ k \end{array} \right ) 
m^{2(N-k)}\langle  \lambda_1^2 \cdots \lambda_k^2\rangle.
\ee
There are $$\left (\begin{array}{c}  k \\ 2l \end{array} \right ) 
(2l-1)!!$$ ways of selecting $l$ pairs from $\lambda_1^2 \cdots \lambda_k^2$.
The average of each pair of different eigenvalues is given by
\be
\langle \lambda_m^2 \lambda_n^2 \rangle =
\sigma^4 +C_2,
\ee
where $\sigma^2$ is the expectation value of $\lambda_k^2$ and $C_2$
is the connected correlator
\be
C_2 = \langle \lambda_m^2 \lambda_n^2 \rangle - 
\langle \lambda_m^2 \rangle \langle \lambda_n^2 \rangle, \qquad m\ne n.
\ee
This results in the partition function
\be
Z(m) = \sum_{k=0}^N \sum_{l=0}^{[\frac k2 ]} m^{2(N-k)} 
\left (\begin{array}{c}  N \\ k \end{array} \right ) 
\left (\begin{array}{c}  k \\ 2l \end{array} \right ) 
(2l-1)!! C_2^l \sigma^{2(k-2l)}.
\ee
After interchanging the two sums, one can easily show that $Z(m)$ can be
expressed as a multiple of a Hermite polynomial
\be
Z(m) = (-\frac{C_2}2)^{N/2} {\rm H}_N((\sigma^2+m^2)/\sqrt{-2C_2}).
\ee
In terms of the zeros of the Hermite polynomials, $z_k$, the zeros of
the partition function are located at
\be
m_k^2 = z_k \sqrt{-2C_2} -\sigma^2.
\ee
Asymptotically, the zeros of the Hermite polynomials are given by
$z_k \approx \pi k/2\sqrt N$ (with integer $\pm k$).
In order for the zeros to join into a cut in the thermodynamic limit, they have
to be spaced as $ \sim 1/N$.  This requires that 
\be
C_2 \sim -\frac 1N.
\ee 
The density of zeros is then given by
\be
\frac{dk}{dm} \sim N m.
\ee
We conclude that pair correlations are sufficient to generate a cut of
$Z(m)$ in the complex $m$-plane, but the chiral symmetry remains unbroken.
Pair correlations alone cannot suppress the effect of the fermion 
determinant.

\subsection{Leutwyler-Smilga Sum Rules}
We have shown that pair-correlations are not sufficient to generate
a discontinuity in the chiral condensate. In this subsection we 
start from the assumption that
chiral symmetry is broken spontaneously, and look for consistency 
conditions that are imposed
on the Dirac spectrum. As has been 
argued by Gasser and Leutwyler\refnote{\cite{GL}} and 
Leutwyler and Smilga\refnote{\cite{LS}}, in the mesoscopic range (\ref{range}),
the  mass dependence of the QCD partition function is given by 
(for simplicity, all quark masses have been taken equal)
\be
Z^{\rm eff}(m,\theta) 
\sim \int_{U\in G/H} dU e^{mV\Sigma {\rm Re\,Tr}\, Ue^{i\theta/N_f}}.
\label{zm}
\ee
The integral is over the Goldstone manifold associated with chiral
symmetry breaking from $G$ to $H$. For three or more colors with
fundamental fermions $G/H = SU(N_f) \times SU(N_f)/SU(N_f)$.
The finite volume 
partition function in the sector of topological charge $\nu$ follows
by Fourier inversion according to (\ref{znum}).
The partition function for $\nu = 0$ is thus given by (\ref{zm})
with the integration over $SU(N_f)$ replaced by an integral over
$U(N_f)$. The case of $N_f= 1$ is particularly simple. 
Then only a $U(1)$ integration remains, and 
the partition function is given by\refnote{\cite{LS}}
$Z_{\nu = 0}^{\rm eff}(m) = I_0(mV\Sigma)$.
Its zeros are regularly spaced along 
the imaginary axis in the complex $m$-plane, and, in the thermodynamic limit, 
they coalesce into a cut.

The Leutwyler-Smilga sum-rules are obtained by expanding the partition
function $Z_\nu(m)$ in powers of $m$ before and after 
averaging over the gauge field configurations and equating the 
coefficients. This corresponds to an expansion in powers of $m$ of
both the QCD partition function (\ref{QCDpart}) and the finite
volume partition function (\ref{zm}) in the sector
of topological charge $\nu$.
As an example, 
we consider the coefficients of $m^2$ in the 
sector with $\nu = 0$. This results in the sum-rule
\be
\langle {\sum}' \frac 1{\lambda_k^2}\rangle  = \frac {\Sigma^2 V^2}{4N_f},
\label{LS2}
\ee
where the prime indicates that the sum is restricted to nonzero positive 
eigenvalues.

The next order sum rules are obtained by equating the coefficients of order 
$m^4$. They can be combined into
\be
\langle {\sum_{k,l}}' \frac 1{\lambda_k^2 \lambda_l^2}\rangle  
-\langle {\sum_k}' \frac 1{\lambda_k^2}\rangle \langle  {\sum_l}' 
\frac 1{\lambda_l^2}
\rangle
= \frac {\Sigma^4 V^4}{16N_f^2(N_f^2 -1)}.
\label{LSpair}
\ee
We conclude that chiral symmetry breaking leads to correlations
of the inverse eigenvalues. However, if one performs an analysis similar 
to the one in previous section, it can be shown easily
that pair correlations given by (\ref{LSpair}) do not result
in a cut in the complex $m$-plane. Apparently, chiral symmetry breaking 
requires a subtle interplay of all types of correlations.  

For two colors with fundamental fermions or for adjoint fermions the
pattern of chiral symmetry breaking is different. 
Sum rules for the inverse eigenvalues can be derived 
along the same lines. The general expression for the simplest sum-rule
can be summarized as\refnote{\cite{SmV,HVeff}}
\be
\langle {\sum}' \frac 1{\lambda_k^2}\rangle  = \frac {\Sigma^2 V^2}
{4(|\nu| +({\rm dim}(G/H) + 1)/N_f)},
\ee
where ${\rm dim}(G/H)$ is the number of generators of the coset manifold
in (\ref{zm}).

The Leutwyler-Smilga sum-rules can be expressed as an integral over the average
spectral density and spectral correlation functions. For the sum rule
(\ref{LS2}) this results in
\be
\frac 1{V^2 \Sigma^2} \int \frac {\langle\rho(\lambda)\rangle
d\lambda}{\lambda^2} = \frac 
1{4N_f}.
\ee
If we introduce the microscopic variable 
\be
u = \lambda V \Sigma,
\ee
this integral can be rewritten as
\be
\int \frac 1{V\Sigma} \langle\rho(\frac u{V\Sigma}) \rangle
\frac {du}{u^2} = \frac 1{4N_f}.
\ee
The thermodynamic limit of the combination
that enters in the integrand, 
\be
\rho_S(u) = \lim_{V\rightarrow \infty} \frac 1{V\Sigma} \langle
\rho(\frac u{V\Sigma})\rangle,
\label{rhosu}
\ee
will be called the {\it microscopic
spectral density}\refnote{\cite{SVR}}. This limit exists if chiral symmetry is
broken. Our
conjecture is that $\rho_S(u)$ is a universal function that only depends
on the global symmetries of the QCD partition function. Because of universality
it can be derived from the simplest theory with the global 
symmetries of the QCD
partition function. Such theory is a chiral Random Matrix Theory which
will be introduced later in these lectures.

We emphasize again that the
$U_A(1)$ symmetry of the QCD Dirac spectrum leads to two 
different types of eigenvalue correlations: spectral
correlations in the bulk of the spectrum and spectral correlations near zero 
virtuality. The simplest example of correlations of the latter type is
the microscopic spectral density defined in (\ref{rhosu}).

We close this subsection with two unrelated side remarks. First,  the QCD 
Dirac operator 
is only determined up to a constant matrix. We can exploit this freedom to 
obtain a Dirac operator that is maximally symmetric. 
For example, the Wilson lattice QCD Dirac operator, $D^W$, is neither Hermitean
nor anti-Hermitean, but $\gamma_5 D^W$ is Hermitean.

Second, the QCD partition function 
can be expanded in powers of $m^2$ 
before or after averaging over the gauge field configurations. In the latter
case one obtains sum rules for the inverse zeros of the partition function.
As an example we quote,
\be
\left . \sum \frac 1{m_k^2} \right |_{\nu = 0}= \frac {\Sigma^2 V^2}4,
\ee
where we have averaged over field configurations with zero topological charge.

\section{SPECTRAL CORRELATIONS IN COMPLEX SYSTEMS}

\subsection{Statistical Analysis of Spectra}
Spectra for a wide range of complex quantum systems 
have been studied both experimentally and numerically (a excellent 
recent review has been given by Guhr, M\"uller-Groeling and 
Weidenm\"uller\refnote{\cite{hdgang}}).
One basic observation
is that the scale of variations of the average spectral
density and the scale of the spectral fluctuations separate. 
This allows us to unfold the spectrum, i.e. we rescale the 
spectrum in units of the local average level spacing. 
Specifically, the unfolded spectrum is given by
\be
\lambda_k^{\rm unf} = \int_{-\infty}^{\lambda_k} \langle \rho(\lambda')\rangle
d \lambda',
\ee
with unfolded spectral density 
\be
\rho_{\rm unf} (\lambda) = \sum_k \delta(\lambda -\lambda_k^{\rm unf}).
\ee

The fluctuations of the
unfolded spectrum can be measured by suitable statistics. We will consider the
nearest neighbor spacing distribution, $P(S)$, and moments of the number of
levels in an interval containing $n$ levels on average. In particular, we
will consider the
number variance, $\Sigma_2(n)$, and the first two cumulants, $\gamma_1(n)$ and
$\gamma_2(n)$. Another useful statistic is the
$\Delta_3(n)$-statistic introduced by Dyson and Mehta\refnote{\cite{deltaDM}}.
It is related to $\Sigma_2(n)$
via a smoothening kernel. The advantage of this statistic is that its 
fluctuations as a function of $n$ are greatly reduced. 
Both $\Sigma_2(n)$ and $\Delta_3(n)$ can be obtained from the pair correlation
function defined as
\be
Y_2(\lambda,\lambda') =-\langle{\rho_{\rm unf}(\lambda)
\rho_{\rm unf}(\lambda')} \rangle+
\langle {\rho_{\rm unf}(\lambda)}\rangle \langle {\rho_{\rm unf}(\lambda')}
\rangle.
\ee

Analytical expressions for the above  statistics can be obtained for
the eigenvalues of the invariant random matrix ensembles. 
They are defined as ensembles of Hermitean 
matrices with Gaussian independently distributed matrix elements, i.e.
with probability distribution given by
\be
P(H) \sim e^{-\frac {N\beta} 2 {\rm Tr } H^\dagger H}.
\label{phinv}
\ee 
Depending on the anti-unitary symmetry, the matrix elements are real, complex
or quaternion real. They are called the Gaussian Orthogonal Ensemble (GOE),
the Gaussian Unitary Ensemble (GUE) and the Gaussian Symplectic Ensemble
(GSE), respectively. Each ensemble is characterized by its Dyson index
$\beta$ which is defined as the number of independent variables per matrix
element. For the GOE, GUE and the GSE we thus have $\beta =1, \, 2$ and 4,
respectively. 

Independent of the value of $\beta$, the average spectral
density is the semicircle,
\be
\langle \rho(\lambda)\rangle = \frac N \pi \sqrt{2-\lambda^2}.
\ee

Analytical results for all spectral correlation functions 
have been derived for each of the three ensembles\refnote{\cite{Mehta}} 
via the orthogonal polynomial method. 
We only quote the most important results.
The nearest neighbor spacing distribution, which is known exactly in terms of a
power series, is well approximated by
\be
P(S) \sim S^{\beta}\exp(-a_\beta S^2),
\ee 
where $a_\beta$ is a constant of order one.
The asymptotic behaviour of the pair correlation function is given
by\refnote{\cite{Mehta}}
\be
Y_2(\lambda,\lambda') &\sim& \frac {1}{\pi^2(\lambda-\lambda')^2} 
\quad {\rm for} \quad \beta = 1,\\
Y_2(\lambda,\lambda') &\sim& \frac {\sin^2\pi(\lambda 
-\lambda')}{\pi^2(\lambda-\lambda')^2} \quad {\rm for} \quad \beta = 2,\\
Y_2(\lambda,\lambda') &\sim& -\frac{\cos 
2\pi(\lambda-\lambda')}{4(\lambda-\lambda')} +\frac {1+ (\pi/2)\sin2\pi(\lambda 
-\lambda')}{4\pi^2(\lambda-\lambda')^2} \quad {\rm for} \quad \beta = 4.
\ee
The $1/(\lambda-\lambda')^2$ tail of the pair correlation function results
in a logarithmic dependence of the asymptotic behavior of $\Sigma_2(n)$
and $\Delta_3(n)$,
\be
\Sigma_2(n) \sim (2/\pi^2\beta) \log n \quad {\rm and} \quad  
\Delta_3(n) \sim \beta \Sigma_2(n)/2.
\ee
Characteristic features of random matrix correlations are
level repulsion at short distances and a strong suppression
of fluctuations at large distances.

For uncorrelated eigenvalues the level repulsion is absent and
one finds 
\be
P(S) = \exp(-S),
\ee
and 
\be
\Sigma_2(n) = n  \quad {\rm and} \quad \Delta_3(n) = n/15.
\ee

\subsection{Spectral Universality}

The main conclusion of numerous studies of eigenvalue spectra 
of complex systems is that
spectral correlations of classically chaotic systems are given 
by RMT\refnote{\cite{hdgang}}. 
As illustration of this so called Bohigas conjecture, 
we mention three examples from completely
different fields. The first example is the nuclear data ensemble in which
the above statistics are evaluated by superimposing level spectra of
many different nuclei\refnote{\cite{Haq}}. The second example concerns 
correlations of acoustic resonances in irregular 
quartz blocks\refnote{\cite{Guhr}}. In both cases the statistics that 
were considered are,
within experimental accuracy, in complete agreement with
the GOE statistics. The third example
pertains to the zeros of Riemann's zeta function. Extensive numerical
calculations\refnote{\cite{Odlyzko}} have shown that asymptotically, for large
imaginary part, the correlations between the zeros are given by the GUE.

The Gaussian random matrix ensembles introduced above
can be obtained\refnote{\cite{Mehta}} from two assumptions:
i) The probability distribution is invariant under unitary
transformations;   ii) The matrix elements are statistically independent.
If the invariance assumption is dropped it can be shown
with the theory of free random 
variables\refnote{\cite{Voiculescu}} 
that the average spectral density is still given by a semicircle
if the variance of the probability distribution is finite. For example,
if the matrix elements are distributed according to a rectangular 
distribution, the average spectral density is a semicircle.
On the other
hand, if the independence assumption is released the average spectral
density is typically not a semicircle. For example, this is the case
if the  quadratic potential in the probability distribution is replaced
by a more complicated polynomial potential $V(H)$. Using the supersymmetric
method for Random Matrix Theory, it was shown by 
Hackenbroich and Weidenm\"uller\refnote{\cite{Hack}} that 
the same supersymmetric
nonlinear $\sigma$-model is obtained for a wide range of potentials $V(H)$.
This implies that spectral
correlations of the unfolded eigenvalues are independent of the
potential. Remarkably, this result could
be proved for all three values of the Dyson index. 

Several examples have been considered where both the invariance assumption
and the independence assumption are relaxed. We mention $H \rightarrow
H+A$, where $A$ is an arbitrary fixed matrix, and
the probability distribution of $H$ is given by a polynomial 
$V(H)$. It was shown
by P. Zinn-Justin\refnote{\cite{Pzinn}} 
that also in this case the spectral correlations are given
by the invariant random matrix ensembles.  For a Gaussian probability
distribution the proof was given 
by Br\'ezin and Hikami\refnote{\cite{Brezin-Hikami}}.

The domain of universality has been extended 
in the direction of real physical systems by means of 
the Gaussian embedded ensembles\refnote{\cite{french,zirntwo}}. The 
simplest example is the ensemble of matrix elements of $n$-particle Slater
determinants of a two-body operator with random two-particle matrix 
elements. It can be shown analytically 
that the average spectral density is a Gaussian.
However, according to substantial numerical evidence, the spectral correlations
are in complete agreement with the invariant random matrix 
ensembles\refnote{\cite{french}}.

A large number of examples have been found that fall into one of
universality classes of the invariant random matrix ensembles. This calls 
out for a more general approach. 
Naturally, one thinks in terms of the renormalization
group. This approach was pioneered by Br\'ezin and 
Zinn-Justin\refnote{\cite{Brezinn}}. The idea is to integrate out rows and 
columns of a random matrix and to show that the Gaussian ensembles
are  a stable fixed point. This was made more explicit in a paper
by Higuchi {\it et al.}\refnote{\cite{Nishigaki-unir}}.
However, much more work is required 
to arrive at a natural proof of spectral universality.

Although the above mentioned universality studies provide support for
the validity of the Bohigas conjecture, the ultimate goal is to derive it
directly from the underlying classical dynamics. 
An important first step in this direction
was made by Berry\refnote{\cite{Berry}}. He showed that the asymptotics 
of the two-point
correlation function is related to sum-rules for isolated
classical trajectories. Another interesting approach was introduced by
Andreev {\it et al.}\refnote{\cite{andreev}} who 
were able to obtain a supersymmetric nonlinear sigma model from spectral
averaging. In this context we also mention the work of Altland and 
Zirnbauer\refnote{\cite{Martinkick}}
who showed that the kicked rotor can be mapped onto a supersymmetric
sigma model.

\section{CHIRAL RANDOM MATRIX THEORY}
\subsection{Introduction of the Model}
In this section we will introduce an instanton 
liquid\refnote{\cite{shurrev,diakonov}} inspired 
chiral RMT for the QCD partition function. 
In the spirit of the invariant random matrix ensembles 
we construct a model for the Dirac
operator with the global symmetries of the QCD partition function as input, but
otherwise  Gaussian random matrix elements. 
The chRMT that obeys these conditions is defined 
by\refnote{\cite{SVR,V,VZ,Vlattice}}
\be
Z_{N_f,\nu}^\beta(m_1,\cdots, m_{N_f}) = 
\int DW \prod_{f= 1}^{N_f} \det({\rm \cal D} +m_f)
e^{-\frac{N \beta}4 {\rm Tr}V(W^\dagger W)},
\label{zrandom}
\ee
where
\be
{\cal D} = \left (\begin{array}{cc} 0 & iW\\
iW^\dagger & 0 \end{array} \right ),
\label{diracop}
\ee
and $W$ is a $n\times m$ matrix with $\nu = |n-m|$ and
$N= n+m$. 
As is the case in QCD, we assume that $\nu$ does not exceed $\sqrt N$, 
so that, to a good approximation, $n = N/2$. The parameter $N$ 
is identified as the dimensionless volume of space time. 
The potential $V$ is defined by
\be
V(\phi) = \sum_{k\ge 1} \frac {a_k}k \phi^k.
\ee
The simplest case is the Gaussian case when $V(\phi) = \Sigma^2\phi$. Below it 
will be shown that the microscopic spectral density is independent of
the higher order terms in this potential provided that the average
spectral density near zero remains nonzero.
The matrix elements of $W$ are either real ($\beta = 1$, chiral
Gaussian Orthogonal Ensemble (chGOE)), complex
($\beta = 2$, chiral Gaussian Unitary Ensemble (chGUE)),
or quaternion real ($\beta = 4$, chiral Gaussian Symplectic Ensemble (chGSE)).
This partition function is invariant under
\be
W \rightarrow U^\dagger W V
\label{inv}
\ee
where the $n\times n$ matrix $U$ and the $m\times m$ matrix $V$ 
are orthogonal matrices for $\beta=1$, unitary matrices
for $\beta = 2$, and symplectic matrices for $\beta = 4$. This invariance
makes it possible to express the partition function in terms of eigenvalues
of $W$ defined by
\be
W = U^\dagger \Lambda V.
\ee
Here, $\Lambda$ is a diagonal matrix with diagonal matrix elements
$\lambda_k \ge 0$. 
In terms of the eigenvalues the partition function (\ref{zrandom}) is given by
\be
Z_{N_f, \nu}^{\beta}(m_1, \cdots, m_{N_f}) = 
\int d\lambda |\Delta(\lambda_k^2)|^\beta
\prod_k \lambda_k^{2N_f+\beta\nu+\beta-1} \prod_f m_f^\nu 
\prod_{f,k}(\lambda_k^2 + m^2_f)
e^{-\frac{N\beta}4 \sum_k V(\lambda_k^2)},\nonumber\\
\label{zeig}
\ee
where the Vandermonde determinant is defined by
\be
\Delta(\lambda_k^2) = \prod_{k<l} (\lambda_k^2-\lambda_l^2).
\label{vandermonde}
\ee

This model reproduces the following symmetries of the QCD partition
function:

\begin{itemize}
\item The $U_A(1)$ symmetry. All nonzero eigenvalues of the random matrix
Dirac operator occur in pairs $\pm \lambda$ or are zero.

\item  The topological structure of the QCD partition function. The 
Dirac matrix has exactly $|\nu|\equiv |n-m|$ zero eigenvalues. This identifies
$\nu$ as the topological sector of the model.

\item The flavor symmetry is the same as in QCD. For $\beta = 2$ it is
$SU(N_f) \times SU(N_f)$, for $\beta = 1$ it is $SU(2N_f)$ and for
$\beta = 4$ it is $SU(N_f)$.

\item The chiral symmetry is broken spontaneously with 
chiral condensate given by
\be                                                      
\Sigma = \lim_{N\rightarrow \infty} {\pi \rho(0)}/N.
\ee
($N$ is interpreted as the (dimensionless) volume of space
time.) The symmetry breaking pattern
is\refnote{\cite{SmV}} $SU(N_f) \times SU(N_f)/SU(N_f)$,
$SU(2N_f)/Sp(N_f)$ and $SU(N_f)/O(N_f)$ for $\beta = 2$, 1 and 4,
respectively, the same as in QCD\refnote{\cite{Shifman-three}}.

\item The anti-unitary symmetries. 
For three or more colors with
fundamental fermions the Dirac operator has no anti-unitary symmetries,
and generically, the matrix elements of the Dirac operator are
complex. The matrix elements $W_{kl}$ of the corresponding random matrix
ensemble are chosen arbitrary complex as well ($\beta =2$).
For $N_c =2$,  the Dirac operator in the fundamental representation
satisfies
\be
[C\tau_2 K, i\gamma D] = 0,
\ee
where $C$ is the charge conjugation matrix and $K$ is the complex conjugation
operator.
Because, $(C\tau_2 K)^2 =1$, the matrix elements of the Dirac operator
can always be chosen real, and the corresponding random matrix ensemble
is defined with real matrix elements ($\beta = 1$).
For two or more colors with gauge fields
in the adjoint representation the anti-unitary symmetry of the Dirac operator
is given by
\be
[CK, i\gamma D] =0.
\ee
Because $(CK)^2 = -1$, it is possible to rearrange the matrix elements of
the Dirac operator into real quaternions. The matrix elements $W_{kl}$ of the
corresponding random matrix ensemble are chosen quaternion real 
$(\beta =4)$.
\end{itemize}

Together with the invariant random matrix ensembles, the chiral ensembles are
part of a larger classification scheme. 
Apart from the random matrix ensembles discussed in this review, this 
classification also includes random matrix models for disordered
super-conductors\refnote{\cite{Altland}}.
As pointed out by 
Zirnbauer\refnote{\cite{class}}, 
all known universality classes 
of Hermitean 
random matrices are tangent to the large classes of 
symmetric spaces in the classification given by Cartan. 
There is a 
one-to-one correspondence between this classification and the 
classification of the large families of Riemannian symmetric 
superspaces\refnote{\cite{class}}.

\subsection{Calculation of the Microscopic Spectral Density}

The joint eigenvalue distribution of the nonzero eigenvalues for zero masses
follows immediately from the partition function (\ref{zeig}).
For $ N_f$ flavors and topological charge $\nu$ the result for arbitrary
potential is given by\refnote{\cite{V}}
\be
\rho_\beta(\lambda_1, \cdots, \lambda_n) = C_{\beta, n}
\prod_{k<l} |\lambda_k^2 -\lambda_l^2|^\beta \prod_{k}
\lambda_k^{2N_f +\beta\nu+\beta -1}
\exp\left ({-\frac{n\beta}{2} \sum_k V(\lambda^2_k)}\right ),
\label{rhototal}
\ee
where $C_{\beta,n}$ is a normalization constant. For $\beta = 2$ the average
spectral density and the spectral correlation functions can be derived
from (\ref{rhototal}) with the help of the orthogonal polynomial 
method\refnote{\cite{Mehta}}. The orthogonal polynomials in the variable
$x = \lambda^2$ are defined by
\be
\int_0^\infty d x x^a
e^{-nV(x)} P^a_n(x) P^a_m(x)= h_n^a \delta_{mn}.
\label{ortho}
\ee
where
\be
a= N_f + |\nu|\quad {\rm for} \quad \beta = 2.
\ee 
In the Gaussian case, $V(\lambda) = \Sigma^2 \lambda$, 
the associated polynomials are
the generalized Laguerre polynomials. That is why this ensemble  
is also known as the
Laguerre ensemble\refnote{\cite{Kahn,BRONK}}. 
The Vandermonde determinant can be rewritten as 
$\Delta(x_i) = \det_{kl}[ x_k^{l-l}]$.
By the addition of linear combinations of rows this determinant can be
expressed in terms of the orthogonal polynomials (\ref{ortho}), 
$\Delta(x_i) \sim \det_{kl}[P_{l-1}^a(x_k)]$. By multiplying 
the two determinants one obtains
\be
\rho_\beta(x_1, \cdots, x_n) = \frac 1{n!}\det K^a(x_k,x_l),
\ee
where the kernel is defined by
\be
K^a(x, y) = \sum_{k=0}^{n-1} \frac 1{h_k^a} P_k^a(x) P_k^a(y)(xy)^{a/2}
e^{-n(V(x)+V(y))/2}.
\label{kernel}
\ee
The average spectral density is obtained by integrating over 
$x_2,\cdots, x_n$,
\be
\rho(x) = n\int \prod_{k=2}^n d x_k \rho(x, x_2, \cdots , x_n)
= K^a(x,x).
\label{rhofromkernel}
\ee
In terms of the original variables, $\lambda = \sqrt x$,  
the spectral density is given by
\be
\rho(\lambda) = 2\lambda K^a(\lambda^2,\lambda^2).
\ee
In the Gaussian case the spectral density is thus given by
\be
\rho(\lambda)= 2\lambda^{2a+1} \Sigma^{2a+2} n^{a+1} 
e^{-n\Sigma^2\lambda^2} \sum_{k= 0}^{n-1}
\frac 1{h_k^a}L_k^a(\lambda^2\Sigma^2 n) L_k^a(\lambda^2\Sigma^2 n), 
\ee
where the $L_k^a$ are the generalized Laguerre polynomials.
With the help of the Christoffel-Darboux formula the sum can be
expressed into the $n^{\rm th}$ order Laguerre polynomial and its 
derivative (with ${L^a_n}'(x) = - L_{n-1}^{a+1}(x)$),
\be
\rho(\lambda) = \frac{2\Sigma n^{-a} n!}{\Gamma( a +n)} 
(\frac z2)^{2a+1}e^{ -z^2/4n}
\left ( L_{n-1}^{a}(\frac{z^2}{4n})L_{n-1}^{a+1}(\frac{z^2}{4n})-
L_{n}^{a}(\frac{z^2}{4n})L_{n-2}^{a+1}(\frac{z^2}{4n})\right ) .
\ee
Here, we introduced the microscopic variable
\be
z = N \Sigma \lambda = 2n \Sigma \lambda,
\ee
and used the explicit expression for the 
normalization of the Laguerre polynomials, $h_k^a = \Gamma(a+k+1)/k!$.
The microscopic limit is defined by 
\be
\lim_{N\rightarrow \infty} \frac 1{N\Sigma} \rho(\frac{\lambda}{N\Sigma})
\ee
and can be evaluated with the help  of the asymptotic relation
\be
\lim_{n \rightarrow \infty} \frac 1{n^\alpha} L_n^\alpha(\frac xn) =
x^{-\frac {\alpha}{2}} J_\alpha(2 \sqrt x),
\ee
where $J_\alpha$ is the ordinary Bessel function of degree $\alpha$.
In order to take the this limit most conveniently, we substitute
the recursion relations
\be
L_{n-1}^{a+1}(z) &=& L_{n-1}^a(z) +L_{n-2}^{a+1}(z),\nonumber \\
L_{n}^{a}(z) &=& L_{n-1}^{a}(z) + L_{n}^{a-1}(z).
\ee
This results  in the microscopic spectral density 
\be
\rho_S(z)  = \frac {z}{2} (J^2_{a}(z) -J_{a+1}(z)
             J_{a-1}(z)).
\label{micro2}
\ee
{}From the asymptotic relation for the Bessel function
\be
J_\nu(z) \sim \left ( \frac 2{\pi z} \right )^{1/2} \cos(z - \frac{\pi}2 \nu-
\frac {\pi}4)\quad {\rm for} \quad z\rightarrow \infty
\ee
we find that
\be
\lim_{x\rightarrow\infty} \rho_S(x) =  \frac 1{\pi}.
\label{rhoasym}
\ee

The result for the average spectral density follows
from the asymptotic properties of the Laguerre polynomials. It given by
the semicircular distribution 
\be
\rho(\lambda) = (n\Sigma^2/\pi)\sqrt{4/\Sigma^2-\lambda^2}.
\ee
Notice that the microscopic limit of the average spectral density coincides with
the asymptotic limit (\ref{rhoasym}) of the microscopic
spectral density.

The two-point correlation function is obtained by integrating the joint
spectral density over all eigenvalues except two. The microscopic limit
can again be expressed in terms of Bessel functions\refnote{\cite{VZ}}.

The spectral density and the two-point 
correlation function were also derived within the framework
of the supersymmetric method of Random Matrix Theory\refnote{\cite{ast}}. 

The calculation of the average spectral density and the spectral
correlations functions for $\beta =1$ and
$\beta = 4$ is much more complicated. However, with the help of 
skew-orthogonal polynomials\refnote{\cite{Dyson-skew,Mehtaskew,nagao}} exact
analytical results for finite $N$ can be obtained as well.

The microscopic spectral density 
for  $SU(2)$ with fundamental fermions ($\beta =1$) 
is given by\refnote{\cite{V2}}
\be
\rho_S(z) = \frac {1}{4} J_{2a+1}(z{}) &+& \frac {1}{2}
\int_0^\infty dw (zw)^{2a+1} \epsilon(z-w)
\left ( \frac 1w \frac d{dw} - \frac 1z \frac d{dz}\right )\nonumber \\
&\times&
\frac{wJ_{2a}(z{})J_{2a-1}( w{})
-zJ_{2a-1}(z{})J_{2a}(w{})}{(zw)^{2a}(z^2-w^2)},
\label{micro1}
\ee
where $a$ is the combination
\be
a = N_f -\frac 12 +\frac {|\nu|}2.
\label{5.8}
\ee
The microscopic spectral density in the
symmetry class with $\beta = 4$ was first calculated by Nagao and 
Forrester\refnote{\cite{nagao-forrester}}. It is given by
\begin{eqnarray}
  \label{eq3}
  \rho_S(z)=2 z^2\int_0^1duu^2\int_0^1dv&&[J_{4a-1}(2uvz)J_{4a}(2uz)
  -vJ_{4a-1}(2uz)J_{4a}(2uvz)]
\label{micro4}
\end{eqnarray}
with $4a=N_f+2|\nu|+1$.  

The spectral correlations in the bulk of the spectrum are given 
by the invariant random matrix ensemble with the same value
of $\beta$. For $\beta =2$ this was already shown three decades ago
by Fox and Kahn\refnote{\cite{Kahn}}. For $\beta =1$ and $\beta =4 $ this
was only proved recently\refnote{\cite{nagao}}.       

\subsection{Duality between Flavor and Topology}

As one can observe from the joint eigenvalue distribution, for $\beta = 2$
the dependence on $N_f$ and $\nu$ enters only through the combination
$N_f + |\nu|$. This allows for the possibility of trading topology for
flavors. In this section we will work out this duality for the
finite volume partition function. This relation completes the proof
of the conjectured expression\refnote{\cite{Sener3}} for the finite 
volume partition function for
different quark masses and topological charge $\nu$.

For $\beta = 2$ the partition function (\ref{zeig}) obeys the relation
\be
\frac{Z_{N_f,\nu}(m_1, \cdots, m_{N_f})}{ \prod_f m_f^\nu}\sim 
Z_{N_f+\nu,0}(m_1, \cdots, m_{N_f},0,\cdots, 0),
\ee
where the argument of the last factor has $\nu$ zeros.

As an example, the simplest nontrivial identity of this type is given by
\be
Z_{1,1}(m) \sim m Z_{2,0}(m,0).
\ee
Let us  prove this identity without relying on Random 
Matrix Theory. 
According to the definition (\ref{zm}) we have
\be 
Z_{1,1}(m) \sim \int d\theta e^{i\theta} e^{mV\Sigma \cos \theta},
\ee
and 
\be
Z_{2,0}(m,0) \sim \int_{U\in U(2)} dU e^{V \Sigma {\rm Re \,Tr} (MU)},
\ee
where $M$ is a diagonal matrix with diagonal elements $m$ and 0.
The integral over $U$ can be performed by diagonalizing $U$ according
to $U = U_1 e^{i\phi_k} U_1^{-1}$, and choosing $U_1$ and $\phi_k$ as new
integration variables. The Jacobian of this transformation is
\be
J \sim \Delta^2(e^{i\phi_k}).
\ee
The integral over $U_1$ can be performed using the Itzykson-Zuber formula.
This results in
\be
Z_{2,0} \sim \int d\phi_1 d\phi_2 \frac{|e^{-i\phi_1} -e^{-i\phi_2}|^2}
{m(\cos \phi_1 - \cos \phi_2)}
(e^{mV\Sigma \cos \phi_1} - e^{mV\Sigma \cos \phi_2}).
\ee
Both terms in the last factor result in the same contribution to the 
integral. Let us consider
only the first term $\sim \exp(mV\Sigma \cos \phi_1)$. Then the integral
over $\phi_2$ has to be defined as a principal value integral. If we use the 
identity
\be
\frac{|e^{-i\phi_1} -e^{-i\phi_2}|^2}
{(\cos \phi_1 - \cos \phi_2)} = 2\left ( \cos \phi_1
-\frac{\sin \phi_1}{\tan ((\phi_1+\phi_2)/2)}\right ),
\ee
the $\phi_2$-integral of the term proportional to $\sin\phi_1$ gives zero 
because of the principal value prescription. The term proportional to $\cos 
\phi_1$ trivially results in
$Z_{1,1}$. We leave it as an exercise
to the reader to generalize this proof to arbitrary $N_f$ and $\nu$.

The group integrals in finite volume partition function (\ref{zm})
were evaluated by Leutwyler and Smilga\refnote{\cite{LS}} for {\it equal}
quark masses. An expression for {\it different} quark masses was obtained
by Jackson {\it  et al.}\refnote{\cite{Sener3}}. The expression could only
be proved for $\nu = 0$. The above duality can be used to relate
a partition function at arbitrary $\nu$ to a partition function at $\nu =0$.
This completes the proof of the conjectured expression for arbitrary 
topological charge.

\section{UNIVERSALITY IN CHIRAL RANDOM MATRIX THEORY}

In the chiral ensembles, two types of universality studies can be performed.
First, the 
universality of correlations in the bulk of the spectrum. 
As discussed above, they are given by the invariant random matrix ensembles.
Second, the universality of the 
microscopic spectral density and the
eigenvalue correlations near zero virtuality. The aim of such studies
is to show that 
microscopic correlations are stable against deformations of the chiral 
ensemble away from the Gaussian probability distribution. 
Recently, a number of universality studies
on microscopic correlations have appeared. They will be
reviewed in this section.

We wish to emphasize that all universality studies for the chiral ensembles
have been performed for
$\beta =2$. The reason is that $\beta =1$ and $\beta=4$ are mathematically 
much more involved. It certainly would be of interest to extend such studies
to these cases as well.

In addition to the analytical studies to be discussed below the 
universality of the microscopic spectral density also follows from
numerical studies of models with the symmetries of the QCD partition function.
In particular, we mention strong
support in favor of universality 
from a different branch of physics, namely from
the theory of universal conductance fluctuations. In that context, the
microscopic spectral density of the eigenvalues of the transmission matrix
was calculated for the Hofstadter\refnote{\cite{HOFSTADTER}}
model, and, to a high degree of accuracy, it agrees with
the random matrix prediction\refnote{\cite{SLEVIN-NAGAO}}.
Other studies deal with a class random matrix models with matrix elements with
a diverging variance.
Also in this case the microscopic spectral density is given
by the universal expressions\refnote{\cite{Kelner}} 
(\ref{micro1}) and (\ref{micro2}). 

The conclusion that emerges from all numerical and analytical 
work on modified chiral random matrix models
is that the microscopic
spectral density  and the correlations near zero virtuality 
exhibit a strong universality that is comparable to the stability 
of microscopic correlations in the bulk of the spectrum.

Of course, QCD is much richer than chiral Random Matrix Theory. 
One question that
should be asked is at what scale (in virtuality) QCD spectral correlations
deviate from RMT. 
This question has been studied by means of instanton liquid
simulations. Indeed, at macroscopic scales, it was 
found that the number variance
shows a linear dependence instead of the logarithmic dependence 
observed at microscopic scales\refnote{\cite{Osborninst}}. 
More work is needed to determine the point where the crossover between
these two regimes takes place.

\subsection{Invariant Deformations of the Gaussian Random Matrix Ensembles}
In a first class of universality studies one considers probability
distributions that maintain unitary invariance. In this case the joint
probability distribution is given by (\ref{rhototal}). In this section
we consider the simplest nontrivial case with only $a_1$ and $a_2$ different
from zero and present the proof of Akemann, Damgaard, Magnea and
Nishigaki\refnote{\cite{Nishigaki-uni,Damgaard}} for this case.
We wish to point out that the general case only leads to minor complications.

This case was first  studied by Br\'ezin, Hikami and 
Zee\refnote{\cite{brezin-hikami-zee}}. 
They showed that the microscopic spectral density is independent 
of $a_2$. A general proof 
valid for arbitrary potential was given by 
Akemann, Damgaard, Magnea and Nishigaki\refnote{\cite{Nishigaki-uni,Damgaard}}. 
The essence of the proof
is a remarkable generalization of the identity for
the Laguerre polynomials,
\be
\lim_{n \rightarrow \infty}  L_n(\frac x n) =
 J_0(2 \sqrt x) \ ,
\label{asym}
\ee
to orthogonal polynomials determined by an arbitrary potential $V$. 
This relation was proved by deriving a differential equation
from the continuum limit of the recursion relation for orthogonal polynomials.
This proof has been extended to the microscopic correlation functions of
all chiral ensembles in a recent work by
Akemann, Damgaard, Magnea and 
Nishigaki\refnote{\cite{Damgaard}}.

 In the normalization
$P_k(0) =1$, the orthogonal polynomials (\ref{ortho})
satisfy the recursion relation
\be
x P_k(x) = -r_k \left [ P_{k+1}(x) - P_k(x) -Q_k
 (P_k(x)-P_{k-1}(x))\right ],
\label{recursion}
\ee
where 
\be
r_k = -\frac{p_k}{p_{k+1}}\quad{\rm  and}\quad 
Q_k =\frac{h_k r_{k-1}}{r_k h_{k-1}}.
\ee
Here,  the coefficient of $x^k$ in $P_k(x)$ is denoted by $p_k$.
The fractions $r_k$ and the normalizations $h_k$ can be determined from the
relations
\be
\int_0^\infty dx n V'(x)e^{-nV(x)} P_k(x) P_k(x) = 
-\int_0^\infty dx \frac d{dx}\left [e^{-nV(x)} P_k(x) P_k(x)\right ] = 
1,\nonumber \\ \\
\int_0^\infty dx n V'(x)e^{-nV(x)} xP_k(x) P_k(x) -h_k - 2kh_k= 
-\int_0^\infty dx \frac d{dx}\left [e^{-nV(x)} xP_k(x) P_k(x)\right ] = 0
\nonumber \\
\ee
For the potential
\be
V(x) = a_1 x + \frac {a_2}2 x^2, 
\ee 
these relations reduce to
\be
&&n a_1 h_k + n a_2 r_k h_k(1 +Q_k)= 1,\nonumber\\ \\
&&n a_1 r_k h_k (1 + Q_k)
+n a_2 r_k^2\left (h_{k+1} + h_k(1+Q_k)^2 + 
h_{k-1}Q_k^2 \right )- h_k -2kh_k = 0.\nonumber \\
\ee
In order to proceed we take the continuum limit of these relations, i.e.
$k\rightarrow \infty$ and $n\rightarrow \infty$ at fixed  $k/n = t$.
With
\be 
r(t) &=& r(k/n) + {\cal O}(1/n) ,\nonumber \\
h(t)&=& \frac 1n h(k/n) +{\cal O}(1/n^2),
\label{conti}
\ee
to leading order in $1/n$, these recursion relations can be rewritten as
\be
a_1 +2a_2 r(t) = \frac 1{h(t)},\\
a_1 r(t) + 3 a_2r^2(t) = t.
\ee
The functions $r(t)$ and $h(t)$ are given by the solution of these
equations. However, we do not need the explicit solution. A more
useful property is that they satisfy a differential equation 
that does not depend on $a_1$ and  $a_2$:
\be
r'(t)h(t) - 2h'(t)r(t) = h^2(t).
\ee
Remarkably, as was shown by Akemann, Damgaard, Magnea and 
Nishigaki\refnote{\cite{Nishigaki-uni,Damgaard}}, 
this relation is valid for any
polynomial potential. This relation is the essential ingredient
in reducing the continuum
limit of the recursion relation (\ref{recursion}) to a Bessel equation.

To leading order in $1/n$ the r.h.s. of the recursion relation (\ref{recursion})
is zero. We 
have to collect the terms of subleading order. In the continuum limit 
we may write
\be
r_{k-1} &=& r(t) - \frac 1n r'(t), \nonumber \\
h_{k-1} &=& \frac 1n h(t) - \frac 1{n^2} h'(t), \nonumber \\
P_{k-1}(x) &=& P(t,x) - \frac 1n \frac d{dt} P(t,x), \quad {\rm etc.}.
\label{subleading}
\ee
This results in the differential equation
\be
n^2 x P(t, x) = -h(t) \frac {d}{dt} \frac {r(t)}{h(t)} \frac {d}{dt} P(t,x).
\ee
We observe that the continuum  limit of the recursion relation exists if 
we take at the same time the microscopic limit with fixed 
$n \sqrt x$.
If we introduce the new variables
\be
v &=& n\lambda= n\sqrt x,\nonumber\\
u(t) &=& \frac {2\sqrt{r(t)}}{h(t)} = \int_0^{t} \frac {dt'}{\sqrt{r(t')}}
\label{udef}
\ee 
the differential equation reduces to the Bessel equation 
\be
\frac 1u \frac {d}{du} u \frac d {du}P(u,v^2) + v^2P(u,v^2) = 0.
\ee
Here, and below we omit the argument $t$ of $u(t)$.
The general solution of this Bessel equation is given by
\be
P(u,v^2) = A J_0(u v) + B N_0(u v)
\ee
The constants are determined by the boundary condition that
the orthogonal polynomials are normalized as $P_n(0) = 1$. This
results in $A = 1$ and $B = 0$.  

The average spectral density follows from (\ref{kernel}) and 
(\ref{rhofromkernel}). For $a = 0$ (we omit the superscript $a$) we obtain
\be
\rho(\lambda) = 2 \lambda e^{-nV(\lambda^2)} \sum_{k=0}^{n-1} \frac 1{h_k}
P_k(\lambda^2) P_k(\lambda^2).
\ee
With the help of the Christoffel-Darboux formula the sum over the polynomials
can be expressed in $P_{n-1}(\lambda^2)$ and its derivative. This results in
\be
\rho(\lambda) = 2 \lambda e^{-nV(\lambda^2)} \frac{r_{n-1}}{h_{n-1}}
\left ( P_{n-1}'(\lambda^2) P_n(\lambda^2) - P_n'(\lambda^2) P_{n-1}(\lambda^2)
\right ),
\label{darboux}
\ee
where the prime denotes differentiation with respect to $\lambda^2$.
The leading order terms contributing to the continuum limit of 
this equation cancel. The
subleading terms follow from the third equation in (\ref{subleading}). In terms
of the new variables we have,
\be
P_{n-1}(\lambda^2) &\sim& P_n(\lambda^2) - \frac 1n \frac d{dt} 
P(u,\lambda^2)=P_n(\lambda^2) - \frac 1n \frac d{dt} 
J_0(u v),\\
P'_n(\lambda^2) &=&\frac d{d\lambda^2} P_n(\lambda^2)\sim 
n^2\frac d{d v^2} J_0(u v).
\ee 
Taking both the continuum limit and the microscopic limit, eq. (\ref{darboux})
reduces to
\be
\rho(\lambda) &=& 2 \lambda n^2\frac{r(t)}{ h(t)}
\left ( - \frac 1{2v} J_0(uv) \frac d{dv}\frac d{dt} J_0(u v)
+  \frac 1{2v} \frac d{dv} J_0(uv)\frac d{dt} J_0(uv) 
\right )\nonumber \\
&=& {\lambda n^2}  \frac{r(t)uu'}{h(t)} 
\left ( - \frac 1{vu} J_0'(u v) J_0(u v)-
J_0''(u v)J_0(u v) +  J_0'(u v) J_0'(u v) \right ) .
\ee
All polynomials that enter in the precursor to this equation
are of the $n$'th order. Therefore, we can put 
$t= 1$. If we also use the relations $J_0'(z) = -J_1(z)$ and $J_1' = -J_1(z)/z
 +J_0(z)$, and the definition of $u$ (see eq. (\ref{udef})), we obtain
\be
\rho(\lambda) = \frac 12 \lambda n^2 u^2(1)
\left (  J_0(u(1) v)J_0(u(1) v) +  
 J_1(u(1) v)
 J_1(u(1) v) \right ) .
\ee
For $\lambda\rightarrow \infty$ the microscopic spectral density should
coincide with $\rho(0)$. This allows us to identify $u(1)$ as
\be
u(1) = \frac{\pi \rho(0)}n.
\ee
For the microscopic spectral density defined by (with $z=N\Sigma \lambda =
 \pi \rho(0) \lambda$, notice that $N= 2n$)
\be
\rho_S(z) = \lim_{N\rightarrow \infty} \frac 1{N \Sigma} \rho (\frac 
z{N\Sigma}) = \lim_{n\rightarrow \infty} \frac 1{\pi \rho(0)} \rho(
\frac z {\pi \rho(0)}),
\ee
we obtain
\be
\rho_S(z) = \frac z2(J_0^2(z)+ J_1^2(z)),
\ee 
in agreement with the result (\ref{micro2}) derived for the Gaussian chiral
ensemble.
This result is independent of the specific shape of the random matrix 
potential. This proves its universality. 

\subsection{Noninvariant Deformations of the Chiral Gaussian Random
Matrix Ensembles}
In a second class of universality studies one considers 
deformations of the Gaussian random matrix
ensemble that violate unitary invariance. In particular, one has 
considered the case where the 
matrix $W$ in (\ref{diracop}) is replaced by
\be 
W\rightarrow W+ A,
\label{weps}
\ee
whereas the probability distribution of $W$ is Gaussian.
Because of the unitary invariance,  the matrix $A$ 
can always be chosen diagonal. 
The simplest case with 
$A=\pi T$ times the identity was
considered by Jackson {\it et al.}\refnote{\cite{Sener1}}. 
This model provides a schematic model
of the chiral phase transition. In this section we will give a detailed
derivation of the average spectral density. The aim is to shown that 
the spectrum changes dramatically with variations of the temperature
parameter. Nevertheless, it could be shown\refnote{\cite{Sener1}} that the 
microscopic spectral density is temperature independent in the broken
phase.

We will show that for large matrices, the average 
resolvent defined by
\be
G(z) = \frac 1{2n}{\rm Tr} \left \langle  \frac 1{z+i\epsilon -{\cal D}}
\right\rangle
\ee
obeys the cubic equation\refnote{\cite{JV,Stephanov1,Sener1}} (the parameter
$\Sigma = 1$ in (\ref{zrandom}))
\be
G^3 -2zG^2 +G(z^2-\pi^2T^2 + 1) - z = 0.
\label{cubic}
\ee
Here, ${\cal D}$ is the ensemble of random matrices 
\be
{\cal D} = \left ( \begin{array}{cc} 0& W+\pi T \\ W^\dagger + \pi T & 0
\end{array} \right ) ,
\label{matrix}
\ee
with probability distribution given by
\be
P(W) = \exp(-n\rm Tr W W^\dagger).
\label{prob}
\ee
The average over $P(W)$ is denoted by the brackets $\langle \cdots \rangle$.
Because the operator (\ref{matrix})
has only a finite support, it is possible to expand the resolvent in a
geometric series in $1/(z-K)$ for $z$ sufficiently large.  Here, $K$ is
the matrix
\be
K = \left ( \begin{array}{cc} 0 & \pi T \\ \pi T & 0  \end{array} \right ) \ .
\ee
One finds by inspection that ${G(z)}$ satisfies
\be
G(z) = {\rm Tr}\frac 1{z-K} + {\rm Tr}\frac 1{z-K}
\left\langle {
\left ( \begin{array}{cc} 0 & W \\W^\dagger & 0 \end{array} \right )
{\cal G}
\left ( \begin{array}{cc} 0 & W \\W^\dagger & 0 \end{array} \right )
{\cal G}}\right\rangle
\ee
where ${\cal G}$ is the matrix
\be
{\cal G} = \left\langle{\frac 1{z-H}} \right\rangle\ .
\ee
It should be clear that ${\cal G}$ is block
diagonal with the block structure
\newcommand{\id}{\bf 1}
\be
{\cal G} = \left (
\begin{array}{cc} g{\id}_n & h{\id}_n \\ h{\id}_n & g{\id}_n
\end{array} \right ),
\ee
where $\id_n$ is the $n\times n$ identity matrix.
Therefore, we find that $G(z) = g$.  The average over $W$ can be carried out
immediately to give
\be
\left\langle{
\left ( \begin{array}{cc} 0 & W \\W^\dagger & 0 \end{array} \right )
{\cal G}
\left ( \begin{array}{cc} 0 & W \\W^\dagger & 0 \end{array} \right )}
\right\rangle
= \frac 1{n}\left (
\begin{array}{cc} g{\id}_n & 0 \\0& g{\id}_n \end{array} \right ) \ .
\ee
This yields the following matrix equation for $g$ and $h$:
\be
\left ( \begin{array}{cc} z & - \pi T \\ - \pi T & z \end{array} \right )
\left ( \begin{array}{cc} g & h\\h & g
\end{array} \right ) = {\bf 1} +
\left ( \begin{array}{cc} g & 0 \\0 & g
\end{array} \right )
\left ( \begin{array}{cc} g& h\\h& g
\end{array} \right ) \ ,
\ee
which leads to the two independent equations
\be
z g - \pi T h & = & 1 +  g^2 \ , \nonumber\\
z h - \pi T g & = &  g h \ .
\ee
Elimination of $h$ yields the equation
\be
z g - \frac{\pi^2 T^2 g}{z-g} =
1 +  g^2 \ .
\ee
Evidently, it can be rewritten as a
cubic equation for $g$. By taking the trace of ${\cal G}$ one obtains the
announced cubic equation (\ref{cubic}) for $G(z)$.
For completeness we mention that this derivation can be 
rewritten\refnote{\cite{nowakblue}} in terms of the so called blues
function\refnote{\cite{zeeblue}}.

The average spectral density given by
\be
\rho(\lambda) = -\frac 1\pi {\rm Im} G(z=\lambda)
\ee 
is a semicircle at $\pi T=0$ and splits into two arcs at $\pi T = 1$. For 
the spectral density at zero one obtains 
$\rho(0) = \sqrt{1-\pi^2 T^2}$, and therefore chiral
symmetry is broken for $\pi T < 1$, and is restored above this temperature.
In spite of this drastic change in average spectral density, it could
be shown\refnote{\cite{Sener1}} with the help of a supersymmetric formulation
of Random Matrix Theory that the microscopic
spectral density does not depend on $T$.

The super-symmetric method in the first paper by 
Jackson $et$ $al.$\refnote{\cite{Sener1}} is not
easily generalizable to higher order correlation functions. A natural
way to proceed is to employ the super-symmetric method introduced by 
Guhr\refnote{\cite{super-thomas}}. In the case of $\beta = 2$ this 
method results in an analytical expression for the kernel determining all 
correlation functions. This approach was followed in two papers, one by
Guhr and Wettig\refnote{\cite{GWu}} and one by 
Jackson {\it et al.}\refnote{\cite{Seneru}}. 
The latter authors studied 
microscopic correlation functions for $A$ in (\ref{weps}) 
proportional to the identity, whereas Guhr and Wettig  considered
an arbitrary diagonal matrix $A$. It was shown that independent
of the matrix $A$, the correlations are given by the 
Bessel kernel\refnote{\cite{Tracy}}. Of course, a necessary condition on
the matrix $A$ is that chiral symmetry is broken.
Guhr and Wettig also showed that correlations in the bulk of the 
spectrum are insensitive to $A$. 

The deformation $W\rightarrow W+A$ with the probability distribution 
for $W$ given by an arbitrary 
invariant potential has not yet been considered. 
We have no doubt that universality proofs along the lines of
methods developed by P. Zinn-Justin\refnote{\cite{Pzinn}}
can be given.  

\section{LATTICE QCD RESULTS}

Recently, the Dirac spectrum in lattice QCD received a good deal of attention.
In particular, the connection between the topology of
field configurations and the spectrum of the Wilson Dirac operator
has been studied in detail\refnote{\cite{Ivanov,Bardeen,Sailer,Lagae}}.
Other studies are related to the connection between the Wilson Dirac
spectrum and the localization properties of the
eigenfunctions\refnote{\cite{Janssen}}.

In this section we will focus ourselves on the spectral correlations of the
lattice QCD Dirac operator. Both 
correlations in the bulk of the spectrum and the microscopic spectral
density will be studied. Consistent with 
universality arguments presented above, we
find that spectral correlations are in complete agreement with
chiral Random Matrix Theory.

\subsection{Correlations in the Bulk of the Spectrum}
 
Recently, Kalkreuter\refnote{\cite{Kalkreuter}}
calculated $all$ eigenvalues of the $N_c = 2$ lattice Dirac
operator both for Kogut-Susskind (KS) fermions and Wilson fermions
for lattices 
as large as $12^4$. For the
Kogut-Susskind Dirac operator, $D^{KS}$, we use the convention
that it is
anti-Hermitean. Because of the Wilson-term, the Wilson Dirac operator, $D^W$,
is neither Hermitean nor anti-Hermitean. Its 
Hermiticity relation is given by
${D^W}^\dagger = \gamma_5 D^W \gamma_5$.
 Therefore, the operator
$\gamma_5 D^W$ is Hermitean. However, it does not anti-commute with $\gamma_5$,
and its eigenvalues do not occur in pairs $\pm \lambda_k$. 
       
In the case of $SU(2)$,
the anti-unitary symmetry of the Kogut-Susskind 
and the Wilson Dirac operator is given by\refnote{\cite{Teper,HV}},
\be
[D^{KS}, \tau_2 K] = 0, \quad {\rm and} \quad 
[\gamma_5 D^W, \gamma_5 CK\tau_2] = 0.
\ee
Because
\be
(\tau_2 K)^2 = -1 ,\quad {\rm and} \quad (\gamma_5 CK\tau_2)^2 =1,
\ee
the matrix elements of the KS Dirac operator can be arranged into 
real quaternions, whereas the Wilson Dirac operator can be expressed
into real matrix elements.
Therefore, we expect that eigenvalue
correlations in the bulk of the spectrum 
are described by the GSE and the GOE, respectively\refnote{\cite{HV}}.
The microscopic correlations for KS fermions are described by the chGSE. However,
the microscopic correlations for Wilson fermions are not described by
the chGOE but rather by the GOE.  
Because of the anti-unitary symmetry, the 
eigenvalues of the KS Dirac operator are subject to the Kramers degeneracy,
i.e. they are double degenerate.

In both cases, the Dirac matrix is tri-diagonalized by
Cullum's and Willoughby's Lanczos procedure\refnote{\cite{cullum}} 
and diagonalized with a standard
QL algorithm. This improved algorithm makes it possible to obtain $all$
eigenvalues. This allows us to test the accuracy of the eigenvalues
by means of sum-rules for the sum of the squares of the eigenvalues
of the lattice Dirac operator. Typically, the numerical error in the
sum rule is of order $10^{-8}$. 

As an example, in Fig. 1 we show a histogram 
of the overall Dirac spectrum for KS fermions at 
$\beta = 2.4$. Results for the spectral correlations are shown in Figs. 2, 3
and 4. The results for KS fermions are for 4 dynamical flavors
with $ma = 0.05$ on a $12^4$ lattice. The results for Wilson fermions were
obtained for two dynamical flavors on a $8^3\times 12$ lattice. For
the values of $\beta$ and $\kappa$ we refer to the labels of the figure.
For $\beta > 2.4$, with our lattice parameters for KS fermions,
the Dirac spectrum near zero virtuality develops a gap. Of course, this is 
an expected feature of the weak coupling domain. 
For small enough values of $\kappa$
the Wilson Dirac spectrum shows a gap at $\lambda = 0$ as well. In the scaling 
domain the value of $\kappa$ is just above the critical value of $\kappa$.
A qualitative description of the Wilson Dirac spectrum can be obtained
with a random matrix model with the structure of the Wilson Dirac
operator\refnote{\cite{Jurkiewicz}}.

The eigenvalue spectrum is unfolded by fitting a second order polynomial to the
integrated spectral density of a stretch of 500-1000 eigenvalues.
The results for  $\Sigma_2(n)$,
$\Delta_3(n)$ and $P(S)$ in Fig. 2 show an impressive agreement with 
RMT predictions. The fluctuations in $\Sigma_2(n)$ are as expected from RMT.
The advantage of $\Delta_3$-statistic is well illustrated by this figure. 
We also investigated\refnote{\cite{HKV}} the $n$ dependence of the
first two cumulants of the number of levels in a stretch of length $n$. Results
presented in Fig. 4 show a perfect agreement with RMT.
Spectra for different values of $\beta$ have been analyzed as well. 
It is probably no
surprise that random matrix correlations are found at stronger couplings.
What is surprising, however, is that even in the weak-coupling 
domain ($\beta =2.8$) the
eigenvalue correlations are in complete agreement with Random
\begin{center}
\begin{figure}[!ht]
\vspace{5mm}
\centering\includegraphics[width=125mm,angle=0]{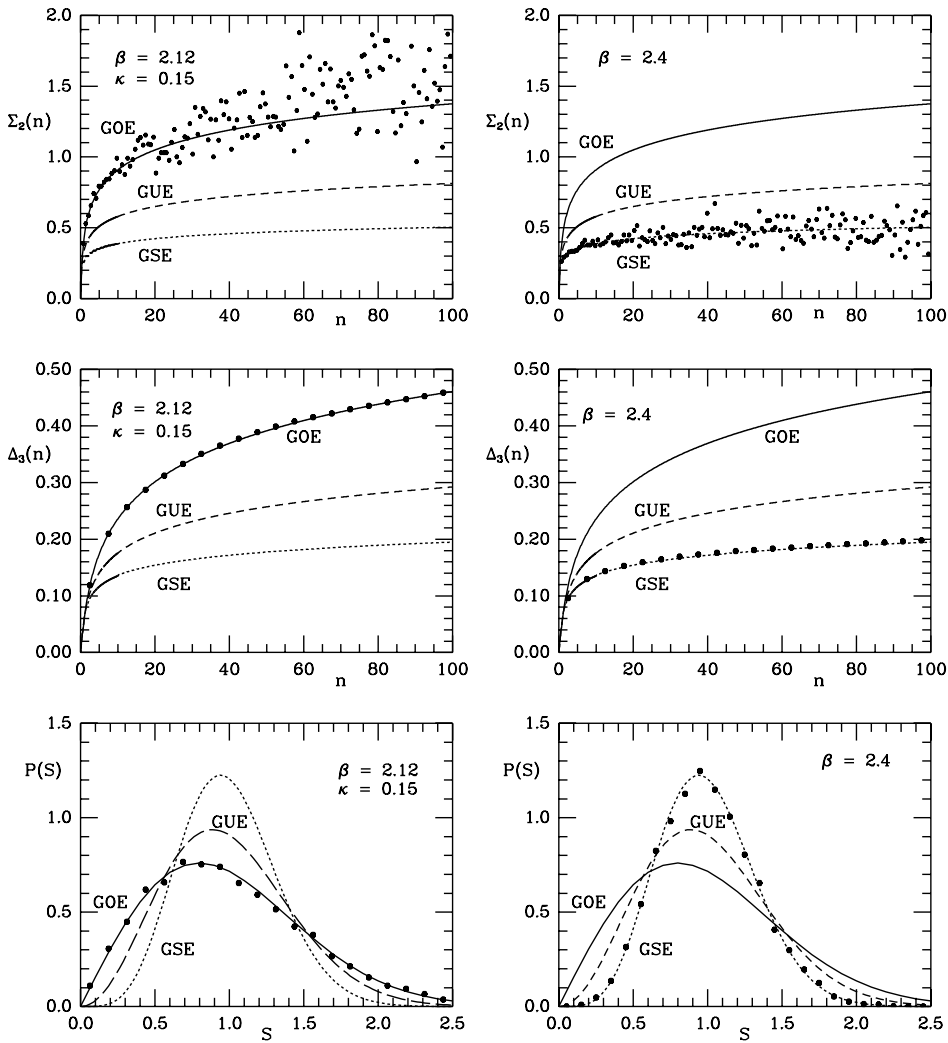}
\caption{
Spectral correlations of Dirac eigenvalues for Wilson fermions
(left) and KS-fermions (right). Results are shown for the number variance,
$\Sigma_2(n)$, the $\Delta_3-$statistic and the nearest neighbor spacing
distribution, $P(S)$. The solid, dashed and dotted curves represent the
analytical result for the GOE, GUE and GSE, respectively.}
\label{fig2}
\end{figure}
\end{center}
Matrix Theory. Finally, we have studied the stationarity of the ensemble
by analyzing level sequences of about 200 eigenvalues (with relatively
low statistics). No deviations from random matrix correlations were observed
all over the spectrum, including the region near  $\lambda = 0$. 
This justifies the spectral averaging which results in
the good statistics in Figs. 2 and 3.

In the case of three or more colors with fundamental fermions, both the 
Wilson and Kogut-Susskind Dirac operator do not possess any anti-unitary 
symmetries. Therefore, our conjecture is that in this case
the spectral correlations in the bulk of the spectrum of 
both types of fermions can be described by
the GUE. In the case of two fundamental colors the continuum theory 
and Wilson fermions are in the same universality class.
It is an interesting question of how spectral correlations of KS fermions
evolve in the approach to the continuum limit. Certainly, the 
Kramers degeneracy of the eigenvalues remains. However, since Kogut-Susskind
fermions represent 4 degenerate flavors in the continuum limit, 
the Dirac eigenvalues should obtain an additional two-fold degeneracy.
We are looking forward to more work in this direction.

\begin{center}
\begin{figure}[!ht]
\vspace{0.3cm}
\centering\includegraphics[width=60mm,angle=0]{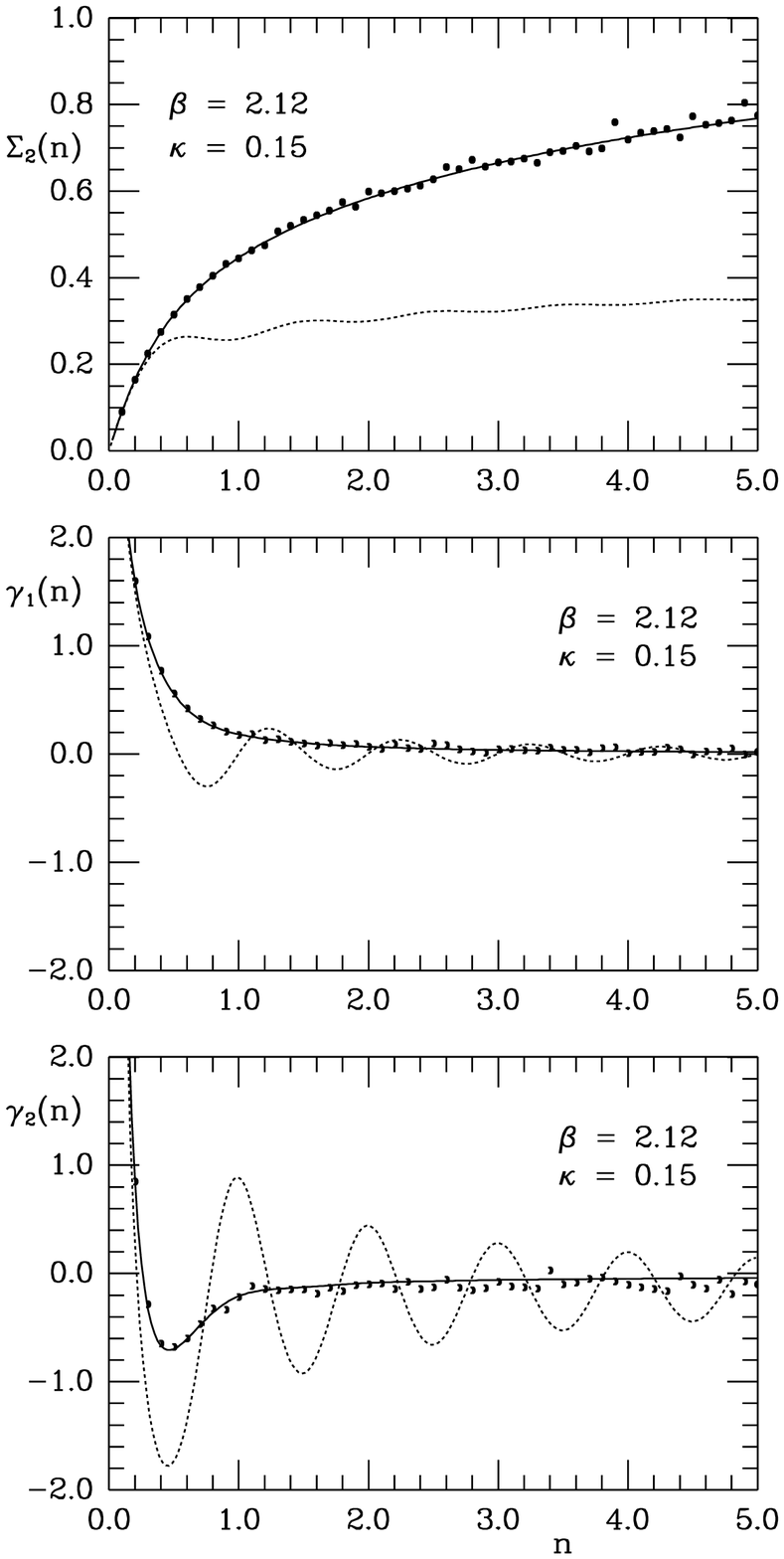}
\includegraphics[width=60mm,angle=0]{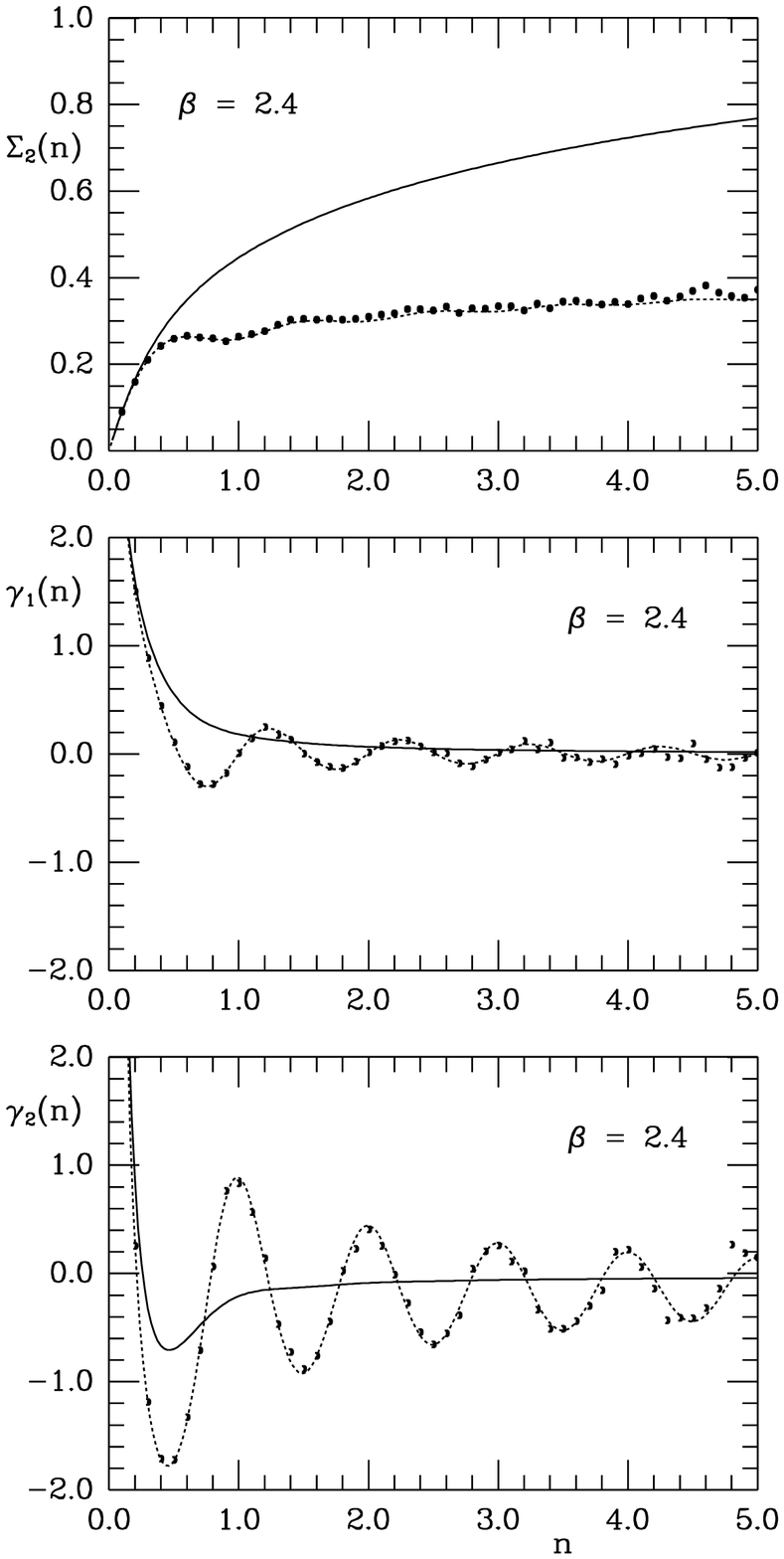}
\caption{The number variance, $\Sigma_2(n)$ and the first two cumulants,
$\gamma_1(n)$ and $\gamma_2(n)$ as a function of $n$ for  eigenvalues
of the Wilson Dirac operator (left) and the Kogut-Susskind Dirac operator 
(right). The full and dotted
curves represent the analytical result for the GOE and the GSE, respectively.}
\label{fig3}
\vspace*{-8mm}
\end{figure}
\end{center}

\subsection{The Microscopic Spectral Density}

The advantage of studying spectral correlations in the bulk of the
spectrum is that one can perform spectral averages instead of ensemble
averages requiring only a relatively small number of equilibrated
configurations. This so called spectral ergodicity cannot
be exploited in the study of the microscopic spectral density.
In order to gather sufficient statistics for the microscopic
spectral density of the lattice Dirac operator
a large number of independent configurations is 
needed. One way to proceed is to generate instanton-liquid configurations
which can be obtained much more cheaply than lattice QCD configurations.
Results of such analysis\refnote{\cite{Vinst}} show that for $N_c=2$
with fundamental fermions the microscopic spectral density is given
by the chGOE. For $N_c =3$ it is given by the chGUE. One could argue that
instanton-liquid configurations can be viewed as 
smoothened lattice QCD configurations. Roughening such configurations
will only improve the agreement with Random Matrix Theory. 

Of course, the ultimate goal is to test the conjecture of microscopic
universality for realistic lattice QCD configurations. In order to obtain a
very large number of independent gauge field configurations 
one is necessarily restricted to relatively
small lattices. The first study in this direction was reported 
recently\refnote{\cite{berbenni,tilohirsch}}. In this work, the 
quenched $SU(2)$ Kogut-Susskind
Dirac operator was diagonalized for lattices with linear dimension of
4, 6, 8 and 10, and a total number of configurations of 9978, 9953, 3896 and
1416, respectively. The results were compared with predictions from the 
chGSE.

\begin{center}
\begin{figure}[!ht]
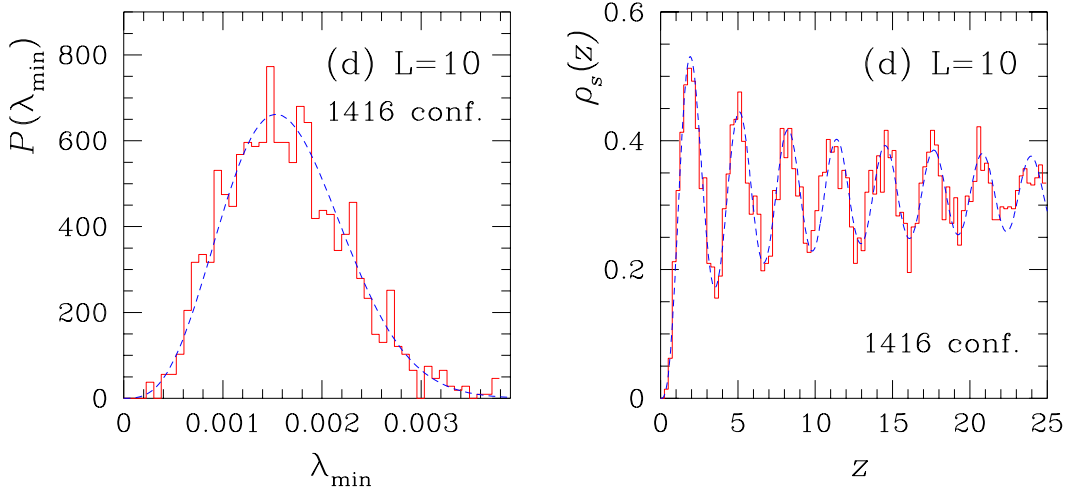

\vspace*{-1cm}
\centering\includegraphics[width=70mm,angle=0]{tilo10small.ps}
\includegraphics[width=70mm,angle=0]{tilo10micro.ps}
\vspace*{-1.5cm}
\caption{The distribution of the smallest eigenvalue (left) and the
microscopic spectral density (right) of the Kogut-Susskind Dirac operator
for two colors and $\beta = 2.0$. Lattice results are represented by
the histogram, and the analytical results for the chGSE are given by the
dashed curves.}
\label{fig4}
\end{figure}
\end{center}
\noindent

\noindent
We only show results for the largest lattice. For more detailed results,
including results for the two-point correlation function,
we refer to the original work. In Fig. 4  we show the distribution of 
the smallest eigenvalue (left) and the microscopic spectral density (right).
The lattice results are given by the full line. The dashed curve represents
the random matrix results. The distribution of the smallest eigenvalue was
derived by Forrester\refnote{\cite{Forrester}} and is given by
\be
P(\lambda_{\rm min}) =\alpha \sqrt{\frac \pi{2}} (\alpha\lambda_{\rm min})^{3/2}
I_{3/2}(\alpha\lambda_{\rm min}) e^{-\frac 12 (\alpha\lambda_{\rm min})^2},
\ee
where $\alpha =V \Sigma$.
The random matrix result for the microscopic spectral density is given
in eq. (\ref{micro4}). We emphasize that the theoretical curves have been
obtained without any fitting of parameters. The input parameter, the
chiral condensate, is derived from the same lattice calculations. 
The above simulations were performed at a relatively strong coupling of
$\beta = 2$. Recently, the same analysis\refnote{\cite{tilopr}} 
was performed for  $\beta = 2.2$
and for $\beta =2.5$ on a $16^4$ lattice. In both cases
agreement with the random matrix predictions was found\refnote{\cite{tilopr}}.

An alternative way to probe the Dirac spectrum is via
the valence quark mass dependence of the chiral 
condensate\refnote{\cite{Christ}} defined as
\be
\Sigma(m) = \frac 1N \int d\lambda \langle\rho(\lambda)\rangle
\frac{2m}{\lambda^2 +m^2}.
\label{sigmam}
\ee
The average spectral density is obtained for a fixed sea quark mass.
For masses well beyond the smallest eigenvalue, $\Sigma(m)$ shows a plateau
approaching the value of the chiral condensate $\Sigma$. 
In the mesoscopic range (\ref{range}), we can introduce $u=\lambda m N$ and
$x= mN\Sigma$ as new variables. Then 
the microscopic spectral density enters in $\Sigma(m)$.
For three fundamental colors the microscopic spectral density
for $\beta =2$ (eq. (\ref{micro2})) applies and the integral over $\lambda$
in (\ref{sigmam}) can be performed analytically. The result is given 
by\refnote{\cite{VPLB}},
\be
\frac {\Sigma(x)}{\Sigma} = x(I_{a}(x)K_{a}(x)
+I_{a+1}(x)K_{a-1}(x)),
\label{val}
\ee
where $a = N_f+|\nu|$ and $I_a$ and $K_a$ are modified Bessel functions.
In Fig. 2 we plot this ratio as a function of $x$ (the 'volume' $V$ is equal
to the total number of Dirac eigenvalues) for
lattice data of two dynamical flavors with  mass $ma = 0.01$ and $N_c= 3$ on a
$16^3 \times 4$ lattice.  We observe
that the lattice data for different values of $\beta$ fall on a single curve.
Moreover, in the mesoscopic range 
this curve coincides with the random matrix prediction for $N_f = \nu = 0$.
\begin{center}
\begin{figure}[!ht]
\centering\includegraphics[width=75mm,angle=270]{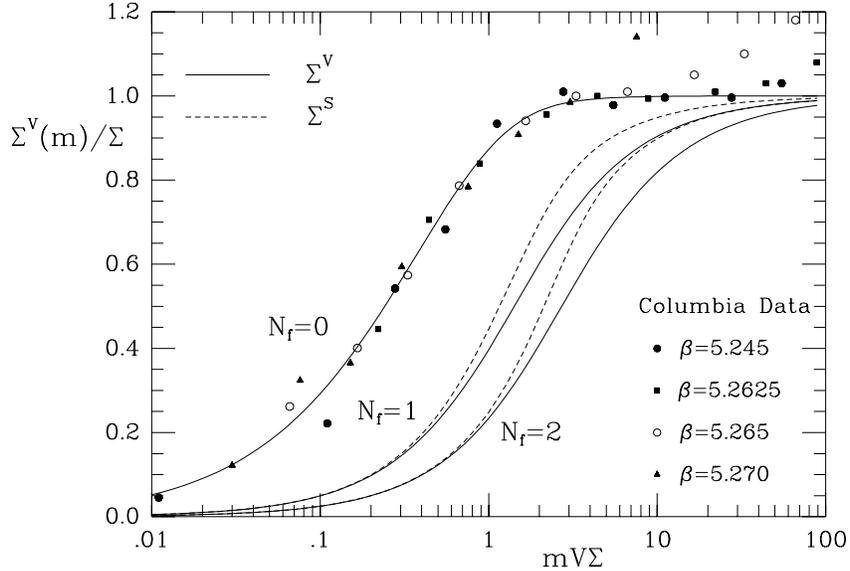}
\caption{The valence quark mass dependence of the chiral condensate
$\Sigma^V(m)$ plotted as $\Sigma^V(m)/\Sigma$ versus $m V\Sigma$. The dots and
squares represent lattice results by the Columbia group\refnote{\cite{Christ}}
 for values of $\beta$ as indicated in the label of the figure.}
\label{fig5}
\end{figure}
\end{center}
Apparently, the zero modes are completely mixed with the much larger number of
nonzero modes. For eigenvalues much smaller than the sea quark mass, one 
expects quenched ($N_f = 0$) eigenvalue correlations. 
In the same figure the dashed curves represent results for the quark mass 
dependence of the
chiral condensate (i.e. the mass dependence for equal valence and sea quark 
masses). 
In the sector of zero topological charge 
one finds\refnote{\cite{jolio,GL,LS}}
\be
\frac{\Sigma^S(u)}{\Sigma} = \frac{I_1(u)}{I_0(u)}\quad {\rm for}\quad
N_f = 1,
\label{sigmam1f}
\ee
and
\be
\frac{\Sigma^S(u)}{\Sigma} = 
\frac{I_1^2(u)}{u(I_0^2(u)-I_1^2(u))}\quad {\rm for}\quad N_f = 2.
\ee
We observe that both expressions do not fit the data. Also notice that, 
according to G\"ockeler {\it  et al.}\refnote{\cite{Gockeler}},
eq. (\ref{sigmam1f}) describes the valence mass dependence of the
chiral condensate for non-compact QED with quenched Kogut-Susskind fermions. 
However, we were not able to derive their result (no derivation is given in 
the paper).

\section{CHIRAL RANDOM MATRIX THEORY AT $\mu \ne  0$}
\subsection{Generalities}
At nonzero temperature $T$ 
and chemical  potential $\mu$ a $schematic$ random matrix 
model of the QCD partition function is obtained by replacing
the Dirac operator in (\ref{zrandom}) 
by\refnote{\cite{JV,Tilo,Stephanov1,Stephanov}}
\be
{\cal D} = \left (\begin{array}{cc} 0 & iW +i\Omega_T +\mu \\
iW^\dagger+i\Omega_T + \mu & 0 \end{array} \right ).
\label{diracmattert}
\ee
Here, $\Omega_T= T \otimes_n(2n+1)\pi {\large\bf 1}$  
are the matrix elements
of $i\gamma_0 \partial_0 $ in a plane wave basis with 
anti-periodic boundary conditions in the time direction.
Below, we will discuss a model with $\Omega_T$ 
absorbed in the random matrix and $\mu \ne 0$. The aim of this
model is to explore the effects of the non-Hermiticity of the 
Dirac operator. For example, 
the random matrix partition function (\ref{zrandom}) 
with the Dirac matrix (\ref{diracmattert}) is well suited for
the study of zeros of this partition function in the complex
mass plane and in the complex chemical potential plane. Numerical results
for the location of zeros could be explained 
analytically\refnote{\cite{Halaszyl}}. 

The term $\mu\gamma_0$ does not affect the anti-unitary
symmetries of the Dirac operator. This is also the case in lattice QCD
where the color matrices in the forward time direction are replaced
by $U \rightarrow e^\mu U$ and in the backward time direction by
$U^\dagger \rightarrow e^{-\mu} U^\dagger$. For this reason the 
universality classes are the same 
as at zero chemical potential. 

The Dirac
operator that will be discussed in this section is thus given by
\be
{\cal D}(\mu) = \left ( \begin{array}{cc} 0 & iW +\mu \\
iW^\dagger+ \mu & 0 \end{array} \right ),
\label{diracmatter}
\ee
where the matrix elements of the $n\times n$ matrix 
$W$ are either real ($\beta = 1$), 
complex ($\beta = 2$) or quaternion real ($\beta = 4$). 
For all three values of $\beta$ the eigenvalues of
${\cal D}(\mu)$ are scattered in the complex plane.

Since many standard random matrix methods 
rely on convergence properties based on the
Hermiticity of the random matrix, direct 
application of most methods is not possible. The simplest way out is
the Hermitization\refnote{\cite{Feinberg-Zee}} 
of the problem, i.e we consider
the Hermitean operator
\be
D^H(z,z^*) = \left ( \begin{array}{cc} \kappa & z-{\cal D}(\mu) \\
z^* - {\cal D}^\dagger(\mu) & \kappa \end{array} \right ).
\label{diracherm}
\ee
For example, the generating function in the supersymmetric method of Random
Matrix Theory\refnote{\cite{Efetov,VWZ}} 
is then given by\refnote{\cite{supernonh,Khoruzhenko,fyodorov,Efetovnh}}
\be
Z(J,J^*) = \left\langle \frac {\det(D^H(z+J,z^*+J^*) }{\det(D^H(z,z^*))} 
\right\rangle.
\label{generator}
\ee
The determinants can be rewritten as fermionic and bosonic
integrals. Convergence is assured by the Hermiticity and by the infinitesimal
increment $\kappa$.
The resolvent follows from the generating function by differentiation
with respect to the source terms
\be
G(z,z^*) = {\rm Tr }\left \langle \frac 1{z-{\cal D}(\mu)} \right\rangle= 
\left . \frac {\partial}{\partial J} Z(J,J^*)\right |_{J=J^* = 0}.
\ee
Notice that, after averaging over the random matrix, the partition function
depends in a non-trivial way on both $z$ and $z^*$. 
The spectral density is then given by
\be
\rho(z) = \frac 1\pi \frac {\partial}{\partial z^*} G(z,z^*).
\ee
Therefore, outside the domain of the eigenvalues $G(z,z^*) $ is a function
of $z$ only, whereas for $z$ inside the domain of eigenvalues, $G(z,z^*)$
should depend on $z^*$ as well.

Alternatively, one can use the replica 
trick\refnote{\cite{Girko,Stephanov}} with
generating function given by
\be
Z(z,z^*,\mu)=\langle {\det}^{N_f} D_H(z,z^*) \rangle.
\label{replica}
\ee 
The resolvent is then given by
\be
G(z,z^*) =  \lim_{N_f \rightarrow 0}\frac 1{2nN_f}
\frac {\partial}{\partial z}\log Z(z,z^*,\mu).
\label{resmu}
\ee
The idea is to perform the calculation for integer values
of $N_f$ and perform the limit $N_f \rightarrow 0$ at the end of the
calculation. 
Although the replica limit, $N_f \rightarrow 0$, fails in 
general\refnote{\cite{martinrep}}, it is expected to work for
$\det D_H(z,z^*)$ because it is positive definite (or zero). Then the
partition function is a smooth function of $N_f$.
For other techniques addressing nonhermitean matrices we refer to the 
recent papers by Feinberg and Zee\refnote{\cite{Feinberg-Zee}} and Nowak and
co-workers\refnote{\cite{Nowak}}.
One recent method that does not rely on the Hermiticity of the random
matrices is the method of complex orthogonal 
polynomials\refnote{\cite{fyodorovpoly}}. This method
was used by Fyodorov {\it et al.}\refnote{\cite{fyodorovpoly}} to
calculate the number variance and the nearest
neighbor spacing distribution in the regime of weakly nonhermitean matrices.
As surprising new result, they found an $S^{5/2}$ repulsion law.
 
In the physically relevant case of QCD with three colors,  
the fermion determinant
is complex for nonzero chemical potential. Its 
phase prevents the convergence of fully unquenched Monte-Carlo
simulations (see Kogut {\it et al.}\refnote{\cite{Klepfish}} for the latest
progress in this direction). However, it is possible to perform quenched
simulations. In such calculations it was found that the critical
chemical potential $\mu_c \sim \sqrt m$, instead of a third of the nucleon
mass\refnote{\cite{everybody}}. This phenomenon was explained
analytically by Stephanov\refnote{\cite{Stephanov}} with the help of the
above random matrix model. He could show that for small $\mu$
the eigenvalues are distributed along the imaginary axis in a band of width
$\sim \mu^2$ leading to a critical chemical potential of $\mu_c \sim \sqrt m$.
A detailed derivation of his solution will be given in the next section.
As has been argued above, the quenched limit 
is necessarily obtained from a partition function in which the 
fermion determinant appears as
\be
\lim_{N_f \rightarrow 0} | \det {\cal D}(\mu)|^{N_f},
\ee
instead of the same expression without the absolute value signs. The
partition function with the absolute value of the determinant can be
interpreted as a partition function of an equal number of fermions and
conjugate fermions. The critical value of the chemical potential, equal
to half the pion mass, is
due to Goldstone bosons with a net baryon number
consisting of a quark and conjugate anti-quark.
The reason that the quenched limit does not correspond to the standard
QCD partition function is closely related to the failure of the replica
trick in the case of a determinant with a nontrivial phase.

\subsection{The Stephanov Solution}

In this section we study the quenched limit of the partition function 
(\ref{replica}). We give a detailed derivation of the results originally
obtained by Stephanov\refnote{\cite{Stephanov}}. 
The determinants in (\ref{replica}) can be written as Grassmann integrals.
Since Grassmann integrals are always convergent
the infinitesimal increment $\kappa$ can be put equal to zero.
This results in the partition function
\be
&&Z(z,z^*, \mu) =
\int {\cal D}W  {\cal D} \psi^* {\cal D} \psi {\cal D} \phi^* {\cal D} \phi
\exp[-{n\Sigma^2} \,{\rm Tr}\, WW^\dagger]\nonumber\\ &\!\times &
\exp \left [- \sum_{k=1}^{N_f}\psi^{k\,*} \left (
\begin{array}{cc} { z} & -iW + \mu \\-i W^\dagger +\mu& { z}
\end{array} \right ) \psi^k
- \sum_{k=1}^{N_f}\phi^{k\,*} \left (
\begin{array}{cc} { z}^* & iW + \mu \\i W^\dagger +\mu& { z}^*
\end{array} \right )\phi^k \right ]
\ ,\nonumber\\
\label{ranpart}
\ee
where $W$ is an arbitrary complex $n\times n$ matrix.
The Gaussian integrals over $W$ can be performed trivially,
\be
Z(z, z^*, \mu) &=&\int {\cal D} \psi^* {\cal D}\psi{\cal D} \phi^* 
{\cal D}\phi \exp \left[ 
{ z}(\psi^{f\,*}_{R\, i} \psi^{f}_{R\, i}
+\psi^{f\,*}_{L\,k}\psi^{f}_{L\, k})
+ { z^*}(\phi^{f\,*}_{R\, i} \phi^{f}_{R\, i}
+\phi^{f\,*}_{L\,k}\phi^{f}_{L\, k})\right .
\nonumber \\ 
&&\hspace{3.8cm} +\mu(\psi^{f\,*}_{R\, i} \psi^{f}_{L\, i}
+\psi^{f\,*}_{L\,k}\psi^{f}_{R\, k}
+\phi^{f\,*}_{R\, i} \phi^{f}_{L\, i}
+\phi^{f\,*}_{L\,k}\phi^{f}_{R\, k}) \nonumber\\
&&\hspace{-1.5cm} -\left . \frac 1{n\Sigma^2 }\left(
\psi^{f\,*}_{L\, k}\psi^{f}_{R\, i} \psi^{g\,*}_{R\,i}\psi^{g}_{L\,k }
-\psi^{f\,*}_{L\, k}\psi^{f}_{R\, i} \phi^{g\,*}_{R\,i}\phi^{g}_{L\,k }
-\phi^{f\,*}_{L\, k}\phi^{f}_{R\, i} \psi^{g\,*}_{R\,i}\psi^{g}_{L\,k }
+\phi^{f\,*}_{L\, k}\phi^{f}_{R\, i} \phi^{g\,*}_{R\,i}\phi^{g}_{L\,k }
\right ) \right ].\nonumber \\
\ee
The four-fermion terms can be
written as the difference of two squares. Each square can be linearized by
the Hubbard-Stratonovitch transformation according to
\be
\exp(-A Q^2) \sim \int d\sigma\exp(-\frac{\sigma^2}{4A} - iQ \sigma) \ \ .
\label{Hubbard}
\ee
Using this, the fermionic integrals can be performed, and the
partition function can be written as an integral over the
complex $2N_f\times 2N_f$ matrices, $\sigma$ and $\tilde \sigma$,
\be
Z(z,z^*, \mu) = \int {\cal D}\tilde \sigma {\cal D} \sigma 
\exp [-{n\Sigma^2} {\rm Tr}(\sigma -\zeta) k (\tilde \sigma-\zeta)k ] 
{\det}^{n} \left ( \sigma \tilde \sigma - \mu^2\right ).
\label{zsteph}
\ee
In a $N_f\times N_f$ block structure notation, the matrices $\sigma,\,
\tilde \sigma,\, k$ and $\zeta$ are given by
\be
\sigma = \left ( \begin{array}{cc} a & ib \\ ic & d \end{array} \right ),
&\qquad& 
\tilde \sigma = \left ( \begin{array}{cc} a^\dagger & ic^\dagger \\ ib^\dagger
 & d^\dagger \end{array} \right ),
\nonumber \\
k = \left ( \begin{array}{cc} 1 & 0 \\ 0 & -1 \end{array} \right ),
&\qquad&
\zeta = \left ( \begin{array}{cc} z & 0 \\ 0 & z^* \end{array} \right ),
\label{matrices}
\ee
where $a, \, b, \, c$ and $d$ are arbitrary complex $N_f \times N_f$ matrices.
The resolvent is obtained by differentiation with respect to $z$ according
to (\ref{resmu}) with the averaged partition function given by (\ref{zsteph}).
This results in
\be
G = \frac 1{2N_f} {\rm Tr} (2 z + a + a^\dagger).
\label{gzmu}
\ee

In the thermodynamic limit the integrals in (\ref{zsteph})
can be evaluated by a saddle-point 
method. Notice that a variable and its complex conjugate have to
be considered as independent integration
variables. The saddle point equations are given by
\be
\frac 1{\tilde \sigma \sigma - \mu^2} \tilde \sigma = k (\tilde \sigma 
+\zeta)k,
\label{m1}\\
\frac 1{\sigma \tilde \sigma - \mu^2} \sigma = k ( \sigma 
+\zeta)k.
\label{m2}
\ee
In general the solution of the saddle point equations is  not unique. However,
a unique solution is obtained from the requirement that $\partial_z^* G(z)$ is
positive definite.
The saddle point equations have two obvious solutions
\be
\sigma = \tilde \sigma \quad {\rm  or } \quad \sigma = k \tilde \sigma k,
\label{general}
\ee
for which the matrix equations (\ref{m1}) and (\ref{m2}) coincide.

It can be shown that the first possibility results in an unphysical solution.
As usual in applications of the replica trick, we assume that the replica
(or flavor) symmetry remains unbroken. This implies that, at the saddle
point, each block in (\ref{matrices}) is diagonal. With (\ref{general})
the solution of our saddle point equations is reduced to a $2\times 2$ matrix
problem
\be
\sigma = (\sigma k \sigma k   -\mu^2)  (k \sigma k + \zeta).
\label{saddle2}
\ee
Consistent with the second solution in (\ref{general}) we 
use the parametrization
\be
\sigma = \left ( \begin{array}{cc} a & ib \\ -ib^* & d \end{array} \right ).
\ee
One  solution of (\ref{saddle2}) can be written down immediately, 
namely $b =b^* = 0$.
Then the equations for $a$ and $d$ reduce to the same cubic equation. 
This solution results in a partition function which factorizes in a product
of a $z$
dependent part and in a $z^*$ dependent part. This resolvent is 
therefore an analytic function of $z$ valid outside the domain of the 
eigenvalues. The cubic
equation for the resolvent in this domain 
can be obtained from the cubic equation (\ref{cubic}) 
by the replacement $T\rightarrow i\mu$.

Let us now focus on the solution inside the domain of eigenvalues with $b\ne 
0$. From the off-diagonal elements of (\ref{saddle2}) we obtain the equations
\be
&&(a+d)(d-z^*)-(a^2-bb^*-\mu^2) = 1,\nonumber\\
&&(a+d)(a-z)-(d^2-bb^*-\mu^2) = 1.
\label{offdiagonal}
\ee
The difference of these equation results in
\be
2(a+d)(d-a)+  (a+d)(z-z^*) = 0,
\ee
with solution given by
\be
a-d = \frac 12 (z-z^*).
\label{difference}
\ee
{}From the sum of the two equations one obtains
\be
b b^*=  1 + \frac 12(a+d)(z+ z^*) -ad- \mu^2.
\label{sum}
\ee
The boundary of the domain of the eigenvalues is given by the set of
points where this solution merges with the factorized solution, i.e. where 
$b = 0$.
With (\ref{difference}) the condition $b= 0$ can be expressed as an equation
for $a$, or by (\ref{gzmu}) ($G =  a-z$ for diagonal $a$) 
an equation for the resolvent. On the boundary of the domain 
the equation for the resolvent is given by
\be
(2G+2x + i y)x -(G+x+iy)(G+x) + 1-\mu^2 = 0.
\label{bounda}
\ee
For $b=0$ the Stephanov solution and the solution of the cubic equation merge
into each other. We thus have a second order phase transition, and 
on the boundary, the second derivative of the free energy corresponding to
(\ref{zsteph}) vanishes.

{}From the diagonal elements of (\ref{saddle2}) one obtains
\be
 a= (a-z)(a^2-bb^*-\mu^2)+bb^*(a+d),\nonumber\\
 d= (d-z^*)(d^2-bb^*-\mu^2)+bb^*(a+d).
\label{diagonal}
\ee
In the first equation we substitute the first equation of (\ref{offdiagonal})
in the first term of the r.h.s. and the expression (\ref{sum}) for $b b^*$
in the second term of the r.h.s.. This results in
\be
d-a +z +(a+d)(\frac 12(z-z^*)(a-d) -\mu^2 + z z^*)=0.
\ee
Together with (\ref{difference}) this results in an equation for $a$ with
solution
\be
a= \frac x{\mu^2 -x^2} +\frac i2 y.
\ee
For the resolvent one obtains
\be
G = a- z=\frac x{\mu^2 -x^2} -x - \frac i2 y.
\label{resfinal}
\ee
This results in the spectral density
\be
\rho(x,y) = \frac 1{4\pi} \left ( \frac{x^2+\mu^2}{(\mu^2-x^2)^2} - 1\right )
\ee
inside the boundary given by 
\be
y^2 = 4 -4\mu^2+ \frac{4x^2}{\mu^2-x^2} - \frac{x^2}{(\mu^2-x^2)^2}.
\label{albound}
\ee
The latter equation has obtained by the substitution of
(\ref{resfinal}) in (\ref{bounda}). Both (\ref{resfinal}) and (\ref{albound})
were first derived by Stephanov\refnote{\cite{Stephanov,mishapc}}. 
The closed curves in Fig. 6 show the boundary of the domain of eigenvalues
given by (\ref{albound}). Numerical results for the eigenvalues are represented
by the dots in the same figure.
 
For small $\mu$, the width of the domain of eigenvalues is $\sim \mu^2$. 
This explains that the  critical value of the chemical potential
is given $\mu_c \sim \sqrt m$. This result explains the quenched results
found in lattice QCD at nonzero chemical potential.

\subsection{Random Matrix Triality at Nonzero Chemical Potential}

In this section, we study the Dirac operator (\ref{diracmatter}) for 
all three values of $\beta$.
Both for $\beta= 1$ and $\beta = 4$ the fermion determinant, 
$\det ({\cal D}(\mu) +m)$, is real. This
is obvious for $\beta = 1$. For $\beta = 4$ the reality follows from the 
identity $q^* = \sigma_2 q \sigma_2$ for a quaternion real element $q$, and
the invariance of a determinant under transposition. 
We thus conclude that quenching works for an even number of flavors.
Consequently, chiral symmetry will be  restored for arbitrarily
small nonzero $\mu$, whereas a condensate of
a quark and a conjugate anti-quark develops. Indeed,
this phenomenon has been observed
in the strong coupling limit of lattice QCD
with two colors\refnote{\cite{Elbio}}.

\begin{figure}[ht]
\vspace{0.5cm}
\centering\includegraphics[width=90mm,angle=-90]{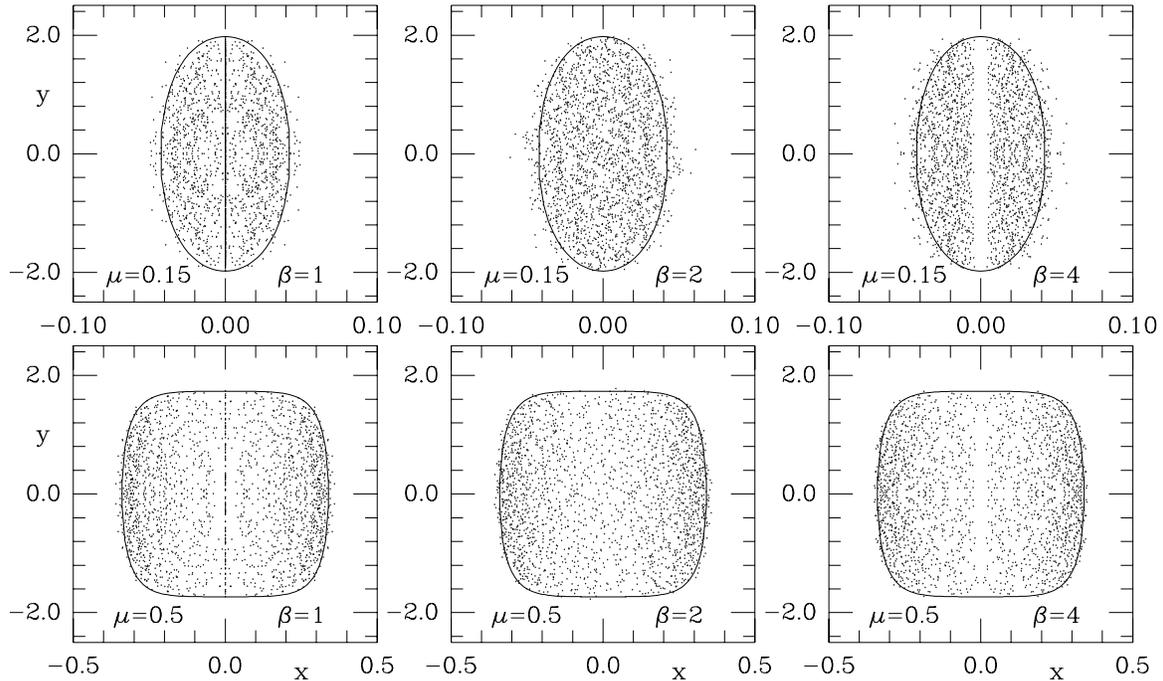}
\caption{
Scatter plot of the real ($x$), and the imaginary
parts ($y$) of the eigenvalues of the random matrix Dirac operator.
The values of $\beta$ and $\mu$ are given in the labels of the figure.
The full curve shows the analytical result for the boundary.}
\label{fig6}
\end{figure}

In the quenched approximation, the spectral properties
of the random matrix ensemble (\ref{diracmatter})
can be easily studied numerically by simply diagonalizing a set of 
matrices with probability distribution (\ref{zrandom}). 
In Fig. 6 we show numerical results\refnote{\cite{Osborn}} 
for the eigenvalues of a few
$100\times  100$ matrices for $\mu = 0.15$ and $\mu = 0.5$.
The dots represent the eigenvalues in the complex plane.
The solid line is the analytical result\refnote{\cite{Stephanov}}
for the boundary of the eigenvalues which is given by
the algebraic curve (\ref{albound}). This result was first derived by
Stephanov for $\beta = 2$ (see previous section). However, 
the method that was used can be extended\refnote{\cite{Osborn}} 
to $\beta = 1$ and $\beta =4$. Although the effective partition function
is much more complicated, it can be shown without too much effort that the
solutions of the saddle point equations are the same 
if the variance of the probability distribution is scaled as $1/\beta$.
In particular, the boundary of the domain of eigenvalues 
is the same in each of the three cases. 
However, as one observes from Fig. 6,  for
$\beta =1$ and $\beta = 4$ the spectral density
deviates significantly from the saddle-point result.
For $\beta = 1$
we find an accumulation of eigenvalues on the imaginary axis, whereas for
$\beta = 4$ we find a depletion of eigenvalues in this domain.
This depletion can be understood as follows. For $\mu = 0$ all eigenvalues
are doubly degenerate. This degeneracy is broken at $\mu\ne 0$ which produces
the observed repulsion of the eigenvalues.

The number of purely imaginary eigenvalues for $\beta = 1$
appears to scale as $\sqrt N$. This
explains that this effect is not visible in a leading order saddle
point analysis. From a perturbative analysis
of (\ref{replica}) one obtains a power series in $1/N$.
Clearly, the $\sqrt N$ dependence
requires a truly nonperturbative analysis
of the partition function (\ref{zrandom}) with 
the Dirac operator (\ref{diracmatter}). 
Such a $\sqrt N$  scaling behavior is typical
for the regime of weak non-hermiticity first identified by Fyodorov
{\it et al.}\refnote{\cite{fyodorov}}. 
Using the supersymmetric method for the generating 
function (\ref{generator}) the $\sqrt N$ dependence was obtained
analytically by Efetov\refnote{\cite{Efetovnh}}.

A similar cut below a
cloud of eigenvalues was found in instanton liquid 
simula\-tions\refnote{\cite{Thomas}}  for $N_c =2$ at $\mu \ne 0$ and
in a random matrix model of arbitrary real 
matrices\refnote{\cite{Khoruzhenko}}. The depletion of the eigenvalues
along the imaginary axis was observed earlier in lattice QCD simulations
with staggered fermions\refnote{\cite{baillie}}. 
Obviously, more work has to be done in order to 
arrive at a complete characterization of 
universal features\refnote{\cite{fyodorovpoly}} in 
the spectrum of nonhermitean matrices.

\section{CONCLUSIONS}
We have argued that there is an intimate relation between correlations of
Dirac eigenvalues and the breaking of chiral symmetry. In the chiral limit,
the fermion determinant suppresses gauge field configurations with 
small Dirac eigenvalues. Correlations counteract this suppression, and
are a necessary ingredient of chiral symmetry breaking. From the study
of eigenvalue correlations in strongly interacting systems, we have
concluded that they are described naturally by a Random Matrix 
Theory with the global symmetries of the physical system. 
In QCD, this
led to the introduction of chiral Random Matrix Theories. 
They provided us with an analytical understanding of the statistical
properties of the eigenvalues on the scale of a typical level spacing.
It could be shown analytically that the microscopic spectral density
is strongly universal. These results constitute the foundation of the
impressive agreement between lattice QCD and chiral Random Matrix
Theory for the microscopic spectral density and for 
spectral correlations in the bulk of the spectrum.

An extension of this model to nonzero chemical potential 
provided us with a complete analytical understanding of the failure of the
quenched approximation observed in lattice QCD simulations at finite density.
Some intriguing properties of 
previously obtained lattice QCD Dirac spectra and instanton liquid Dirac 
spectra at finite density could be explained as well. 
\subsection{Acknowledgements}

This work was partially supported by the US DOE grant
DE-FG-88ER40388.  NATO is acknowledged for financial support. 
In particular, we wish to thank Pierre van Baal for organizing
this wonderful summer school. We benefitted from discussions with 
M. Stephanov and T. Wettig.
M.K. \c Sener and M. Stephanov are thanked for a critical reading 
of the manuscript.
Finally, I thank all my collaborators on whose work this review is based.

\end{document}